%% file: icme2026_template_paper.tex
\documentclass[conference]{IEEEtran}
\IEEEoverridecommandlockouts
\usepackage{cite}
\usepackage{amsmath,amssymb,amsfonts}
\usepackage{algorithmic}
\usepackage{graphicx}
\usepackage{textcomp}
\usepackage{xcolor}
\usepackage{algorithm}
\usepackage{algorithmic}
\usepackage{multirow}
\usepackage{newfloat}
\usepackage{listings}
\usepackage{enumitem} 
\usepackage{arydshln} 
\usepackage{amsmath}
\usepackage{tcolorbox}
\usepackage{xcolor} 
\usepackage{cancel}
\usepackage{caption}
\usepackage{tikz}
\usepackage{url}
\tcbuselibrary{breakable} 
\usepackage{listings} 
\usepackage{graphicx}
\usepackage{booktabs}
\usepackage{algorithm}
\usepackage{algorithmic}
\usepackage{multirow}
\usepackage[normalem]{ulem}
\newcommand{\strike}[1]{\sout{#1}}
\usepackage{newfloat}
\usepackage{listings}
\usepackage{enumitem} 
\usepackage{arydshln} 
\usepackage{amsmath}
\usepackage{tcolorbox}
\usepackage{xcolor} 
\usepackage{cancel}
\usepackage{caption}
\usepackage{tikz}
\tcbuselibrary{breakable} 
\usepackage{listings} 
\usepackage{graphicx}
\usepackage{booktabs}
\newenvironment{promptbox}
{
    \begin{tcolorbox}[
        breakable,
        colback=gray!5, 
        colframe=gray!30, 
        arc=0pt, 
        left=6pt, 
        right=6pt, 
        top=6pt, 
        bottom=6pt, 
        boxrule=0.5pt, 
        leftrule=4pt, 
        before skip=10pt, 
        after skip=10pt, 
    ]
}
{
    \end{tcolorbox}
}
\def\BibTeX{{\rm B\kern-.05em{\sc i\kern-.025em b}\kern-.08em
    T\kern-.1667em\lower.7ex\hbox{E}\kern-.125emX}}
\begin{document}

\title{Towards Automatic Soccer Commentary Generation with Knowledge-Enhanced Visual Reasoning}

\author{
\IEEEauthorblockN{Zeyu Jin\textsuperscript{1,2}, 
Xiaoyu Qin\textsuperscript{1,*}\thanks{*Corresponding authors: Xiaoyu Qin and Jia Jia.}, 
Songtao Zhou\textsuperscript{1}, 
Kaifeng Yun\textsuperscript{1}, 
Jia Jia\textsuperscript{1,2,*}}
\IEEEauthorblockA{\textsuperscript{1}Department of Computer Science and Technology, Tsinghua University,
\textsuperscript{2}BNRist, Tsinghua University\\
\{jinzeyu23, zst24, ykf21\}@mails.tsinghua.edu.cn, \{xyqin, jjia\}@tsinghua.edu.cn}
}

\maketitle
\begin{abstract}
\input{paper/content/0-abstract}

\end{abstract}

\section{Introduction}
\label{sect:introduction}
\input{paper/content/1-introduction}

\section{Related~Works}
\label{sec:relatedworks}
\input{paper/content/2-relatedworks}

\section{Problem~Formulation}
\label{sec:problem-formulation}
\input{paper/content/problem-formulation}

\section{Stage I: Align~Entity~with~Visual~Reasoning}
\label{sec:vidual_reasoning}
\input{paper/content/3-guess-player}

\section{Stage II: Refine~Commentary~with~Knowledge}
\label{sec:knowledge_refinement}
\input{paper/content/4-knowledge-enhanced}

\section{Experiments}
\label{sec:experiment}
\input{paper/content/5-experiments}

\section{Conclusion}
\label{sect:conclusion}
\input{paper/content/6-conclusion}

\section{Acknowledgments}
This work is supported by the National Natural Science Foundation of China Nos. 62425604 and 62502256. It is also supported by Beijing Natural Science Foundation (L257006), High Performance Computing Center, Tsinghua University and Beijing Zitiao Network Technology Co., Ltd.

\bibliographystyle{IEEEbib}
\bibliography{icme2026references}

\newpage
\appendices
\input{paper/content/7-appendix}

\end{document}

%% file: paper/content/0-abstract.tex
Soccer commentary plays a crucial role in enhancing the soccer game viewing experience for audiences.
Previous studies in automatic soccer commentary generation
typically adopt an end-to-end method to generate anonymous \textit{live text commentary}.
Such generated commentary is insufficient in the context of real-world \textit{live televised commentary}, as it contains anonymous entities, context-dependent errors and lacks statistical insights of the game events.
To bridge the gap, we propose \textsc{GameSight}, \textit{a two-stage model to address soccer commentary generation as a knowledge-enhanced visual reasoning task}, enabling live-televised-like knowledgeable commentary with accurate reference to entities (players and teams). 
\textsc{GameSight} starts by performing visual reasoning to align anonymous entities with fine-grained visual and contextual analysis.
Subsequently, the entity-aligned commentary is refined with knowledge by incorporating external historical statistics and iteratively updated internal game state information.
Consequently, 
\textsc{GameSight} improves the player alignment accuracy by 18.5\% on SN-Caption-test-align dataset compared to Gemini 2.5-pro. 
Combined with further knowledge enhancement, \textsc{GameSight} outperforms in segment-level accuracy and commentary quality, as well as game-level contextual relevance and structural composition.
We believe that our work paves the way for a more informative and engaging human-centric experience with the AI sports application.
Demo page: \url{https://gamesight2025.github.io/gamesight2025}.

%% file: paper/content/1-introduction.tex

The soccer industry holds a significant position in the global sports market, with 
a fan base of over five billion people~\cite{fifa_football_landscape_2021}.
As one of the key elements of the viewing experience, soccer commentary is of vital importance in enhancing fan engagement and delivering captivating information.
This has sparked considerable interest in automatic soccer commentary generation~\cite{mkhallati_soccernet-caption_2023,midoglu_mmsys22_2022}. 
Given a segment of soccer game video, end-to-end soccer commentary generation models, e.g., MatchVision~\cite{rao_matchtime_2024} and SoccerComment~\cite{li_multi-modal_nodate}, 
primarily focus on producing descriptions in the style of anonymous \textit{live text commentary}.
However, they fall short when it comes to replicating the real-world \textit{live televised commentary}.

\begin{figure}[h] 
  \vspace{-0.5mm}
  \includegraphics[width=\linewidth]{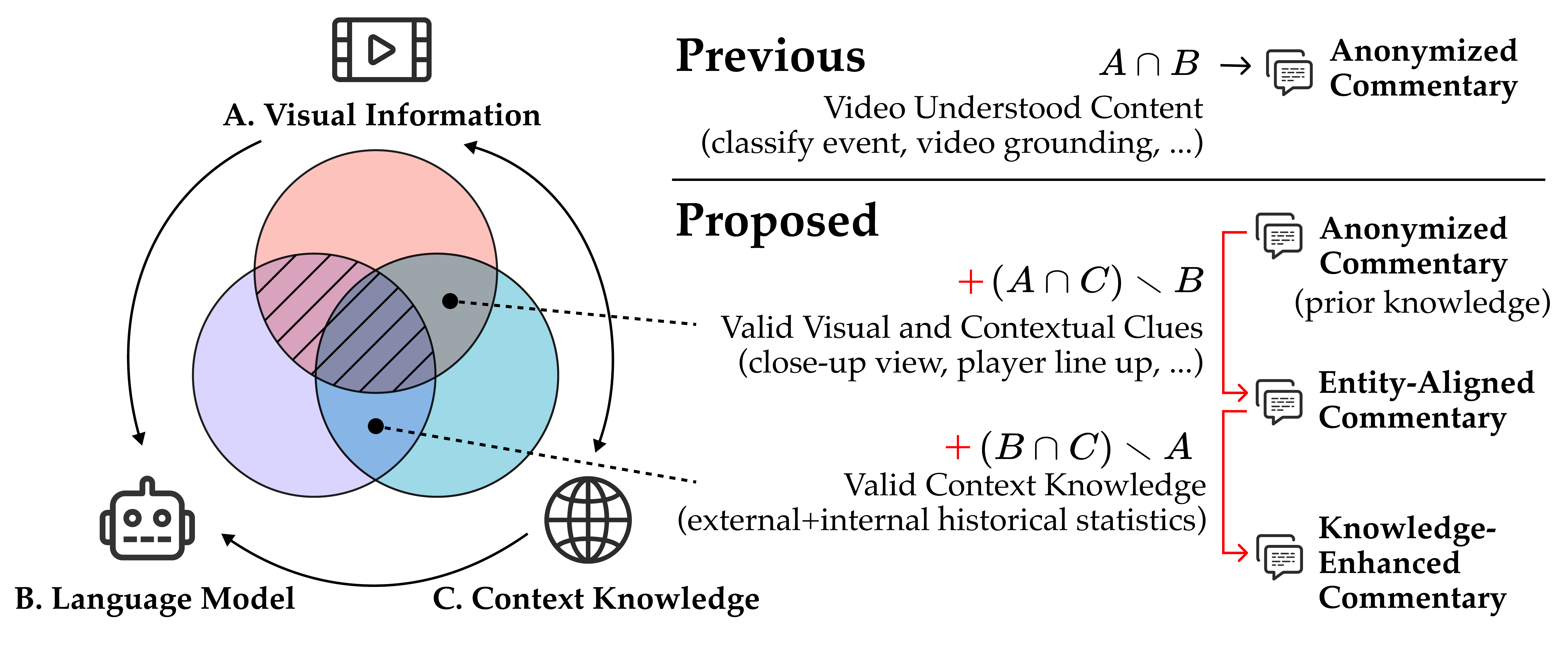}
  \caption{Overview of the proposed \textsc{GameSight}. It incorporates visual and contextual information to generate soccer commentary that is entity-aligned and knowledge-enhanced.}
  \label{fig:venn}
  \vspace{-2mm}
\end{figure}

\begin{figure*}[h]
  \includegraphics[width=\linewidth]{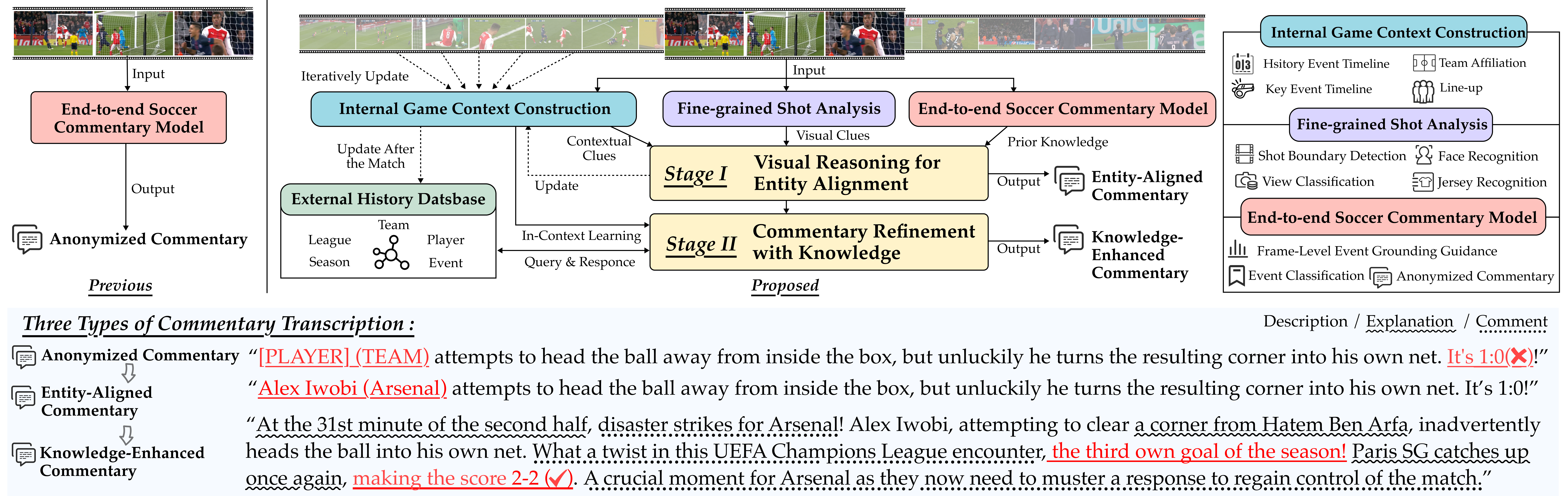}
  \vspace{-2mm}
  \caption{Unlike previous end-to-end models, \textsc{GameSight} is a two-stage system that addresses soccer commentary generation as a knowledge-enhanced visual reasoning task. It revises and enriches the anonymized commentary with \textit{Stage I}: visual reasoning for entity alignment, and \textit{Stage II}: commentary refinement with knowledge.}
  \vspace{-2mm}
  \label{fig:teaser}
\end{figure*}

Professional live televised commentary is characterized by several key elements to captivate the audience: 
\textbf{(1) Accurate reference to entities~\cite{qi_goal_2023}.} 
By aligning names of players and teams in the sportscasting footage and the ongoing events, game commentators help the audience to understand the unfolding action.
\textbf{(2) Awareness of the internal game context.} 
Commentators also integrate the dynamic match context to offer a comprehensive interpretation of current game state. The commentary paragraph should flow continuously, rather than being fragmented, to maintain a cohesive narrative.
\textbf{(3) Utilization of external statistics}. 
Commentary should consist of not only description, but also explanation and comment~\cite{zhang2017shanghai}. Combining historical statistical knowledge serves as the foundation of these broader perspective and deeper insights~\cite{hedrick2000art}.

Previous soccer commentary models fail to meet the above from the following aspects. 
\textbf{Prob.~1: Absence of entity grounding.} Previous models generate lines with anonymized player and team labels as placeholders~\cite{mkhallati_soccernet-caption_2023},
incompetent for aligning the specific players involved in the events.
\textbf{Prob.~2: Context-dependent errors.} 
These models are unaware of broader context as they aim solely at increasing the text similarity in video captioning. 
This leads to context-dependent errors in critical game information, such as incorrect score announcements following goal events.
\textbf{Prob.~3: Lack of statistical insight.} Live text commentary inherently focus on descriptive commentary due to the absence of video, which differs from live televised commentary. As a result, previous models lack the incorporation of statistical knowledge. 

To tackle the three key discrepancies between end-to-end soccer commentary models and human-centric televised commentary, we propose \textsc{GameSight}, a two-stage model that addresses soccer commentary generation as a knowledge-enhanced visual reasoning task.
We move beyond the limited end-to-end captioning by decoupling the task into visual entity grounding and knowledge-driven refinement. This design mimics the cognitive process of human commentators: first identifying the participants through visual and contextual cues, and then enriching the narrative with deep domain knowledge, as illustrated in Fig.~\ref{fig:venn}. By integrating both internal game context and external historical statistics, \textsc{GameSight} produces commentary that is not only visually accurate but also contextually profound.

In summary, our contributions are three-fold:
\begin{itemize}
\item We introduce soccer commentary generation as a knowledge-enhanced visual reasoning task, bridging the gap between automated live-text like commentary and human-centric live televised like commentary.

\item We propose \textsc{GameSight}, a two-stage system that leverages fine-grained visual and contextual information to conduct entity alignment and knowledge enhancement for commentary.

\item Experiments show that 
\textsc{GameSight} outperforms in both segment-level and game-level to generate accurate, informative and engaging commentary, showcasing the capability of AI-based method to enrich the viewing experience in sports analysis.
\end{itemize}





%% file: paper/content/2-relatedworks.tex
\subsection{Automatic Soccer Game Analysis}
Automatic soccer game analysis aimed at reconstructing detailed aspects of a game with two major intentions, to recognize the \textbf{players} or detect the \textbf{events}.

On the player recognition front, jersey number recognition~\cite{Cioppa2022Scaling,balaji_jersey_2023} and game state reconstruction~\cite{Somers2024SoccerNetGameState} have achieved significant progress. However, they usually
rely on high quality, single full-view camera footage and individual player tracking~\cite{cioppa2022soccernet} for identification.

Event classification methods~\cite{kompatsiaris_soccer_2019,sorano_automatic_2020} employ complex pipelines with decomposed subtasks, for example, shot boundary detection~\cite{rao_unified_2020} and play-break segmentation~\cite{tjondronegoro2003sports}. 
Soccer commentary generation represents a higher-level abstraction of event classification to further describe the scene.
SoccerNet-Caption~\cite{mkhallati_soccernet-caption_2023} firstly introduced anonymized soccer commentary, which has become the standard task format~\cite{rao_matchtime_2024,li_multi-modal_nodate}.
MatchVision~\cite{rao_towards_2024} proposed a unified framework with a pretrained spatiotemporal visual encoder.
TimeSoccer~\cite{you2025timesoccerendtoendmultimodallarge} enables game-level commentary generation with global context modeling.

Despite significant progress in commentary generation, there remain limited efforts in \textbf{aligning} player with event. Goal~\cite{qi_goal_2023} 
generates commentary with knowledge triples. However, it suffers from entity mismatching and knowledge deficiency 
In this paper, we perform player alignment from anonymous commentary via a two-stage knowledge-enhanced visual reasoning framework. Our method achieves high entity alignment accuracy and effective incorporation of professional domain knowledge.


\subsection{Reasoning with Video-LMMs}

\textit{Video large multi-modal models} (Video-LMMs)~\cite{li2023videochat, wang2025vcass} have shown remarkable performance in zero-shot video understanding, video grounding, optical character recognition, and event captioning with timestamps. 
These capabilities make them inherently suited for addressing various subtasks in soccer video analysis. 
Despite their potential, Video-LMMs still face challenges in reasoning within multi-modal contexts~\cite{khattak2024complex,jin2026from}, particularly when it comes to integrating knowledge from specialized domains like soccer. 
Related cross-modal alignment work, such as HarmoniVox~\cite{zhou2025harmoniVox}, studies audiovisual harmony, but does not address soccer-specific temporal reasoning or entity-aligned commentary generation.
The domain adaptation of VLM \cite{jiang2025domainadaptationvlmsoccer} excels in understanding soccer concept in 2-second video clips, but it lacks the contextual reasoning capabilities for game-level entity-aligned commentary generation.
In this work, we not only highlight the potential of Video-LMMs for fine-grained sports analysis, but also showcased the effectiveness of supervised fine-tuning (SFT)~\cite{ouyang2022training} and group relative policy optimization (GRPO)~\cite{shao2024deepseekmath} in enhancing the reasoning abilities in such kind of complex, practical multi-model reasoning task.


\subsection{Knowledge-Augmented Generation}


Recent efforts, such as K-SportsSum~\cite{wang2021knowledgeenhancedsportsgame},  SoccerRAG~\cite{strand_soccerrag_2024} and SoccerAgent~\cite{rao2025socceragent}, have introduced knowledge-enhanced techniques in sports domain. However, these works primarily focus on static player attributes or video question answering, leaving the integration of dynamic contextual knowledge in real-world sports applications underexplored. 
Unlike the Retrieval-Augmented Generation (RAG)~\cite{jin2024bider} using retrieval-based models to retrieval information from large text corpus, Knowledge-Augmented Generation (KAG)~\cite{liang2024kag} uses explicitly structured and factual knowledge to inform responses.
By extending the SoccerRAG to SoccerKAG and incorporating dynamic knowledge, we provide LLM with precise knowledge to enable insightful commentary.

%% file: paper/content/problem-formulation.tex

To bridge the gap, we formulate soccer commentary generation as a \textbf{knowledge-enhanced visual reasoning task}. Given a soccer video segment $\mathcal{S}$ and its corresponding initial anonymized commentary $C_A$ (typically generated by an end-to-end model), our goal is to produce a final commentary $C_{KE}$ that is both entity-accurate and contextually insightful.


\subsection{The Two-Stage Design Rationale}
Unlike previous end-to-end approaches, we decouple the task into two specialized stages (as shown in Fig.~\ref{fig:teaser}) to address the three problems:

\textbf{Stage I: Align~Entity~with~Visual~Reasoning (Prob. 1).}
As shown in the pre-experiment (Sec.~\ref{Human Baseline}), human rely on a rich set of contextual and visual cues to understand the event-related players besides the player tracking ~\cite{Somers2024SoccerNetGameState} in long view scenes.
Inspired by this insight, Stage I is trained with \textit{supervised fine-tuning}~(SFT) and \textit{group relative policy optimization}~(GRPO) to combine internal game context and fine-grained shot analysis to transform $C_A$ into entity-aligned commentary $C_{EA}$~(Sec.~\ref{Answer-Guided Reasoning}).
    
\textbf{Stage II: Commentary Refinement with Knowledge (Prob. 2 \& 3).}
An analysis of live televised commentators' practices (Sec.~\ref{Knowledge Reference in Broadcast Commentary}) reveals their frequent and flexible application of relevant knowledge.
In line with this behavior, we develop a soccer \textit{knowledge-augmented generation} (KAG) system and an iteratively updated database of internal game context knowledge in Stage II to generate $C_{KE}$.

\subsection{Formal Objectives}
Formally, the pipeline is modeled as two successive stages:

\textbf{Stage I} (Sec III): $f_{align}(\mathcal{S}, C_A, Context) \rightarrow C_{EA}$. This function focuses on maximizing the grounding accuracy of $C_{EA}$ relative to the ground-truth players in $\mathcal{S}$. The $Context$ includes contextual clues $C_{GS}$~(Sec.~\ref{Context Game State Reconstruction}), visual clues $Scene$~(Sec.~\ref{Fine-grained Visual Details Extraction}), and the prior knowledge guidance $\mathcal{W}$~(Sec.~\ref{Frame-Level Event Grounding Guidance}).

\textbf{Stage II} (Sec IV): $f_{enrich}(C_{EA}, \mathcal{K}_{ext}, \mathcal{K}_{int}) \rightarrow C_{KE}$. This function focuses on generating $\mathcal{C}_{KE}$ with not only statistically accurate and contextually related commentary, but also insightful explanations and comments in a training-free manner, where $\mathcal{K}_{ext}$~(Sec.~\ref{External Soccer RAG System Modification}) and $\mathcal{K}_{int}$~(Sec.~\ref{Internal Game Context Knowledge Construction}) denote external statistics and internal match history, respectively.

%% file: paper/content/3-guess-player.tex
\begin{figure*}[h]
  \vspace{-0.25cm}
  \includegraphics[width=\linewidth]{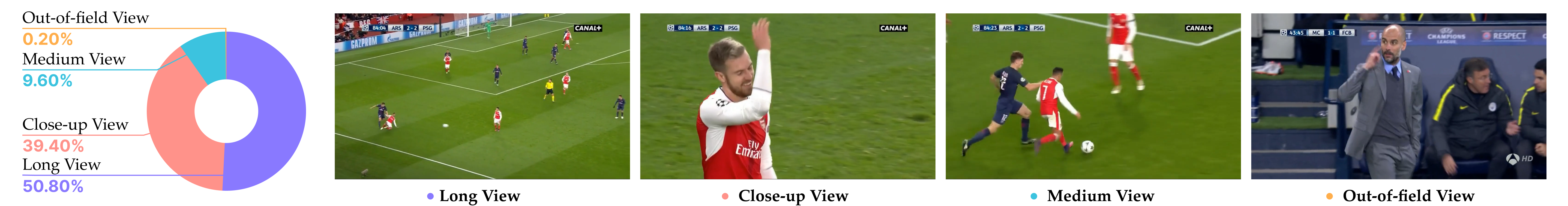}
  \vspace{-0.4cm}
  \caption{Soccer game view shots distribution in time length and view shots examples.}
  \label{fig:view shots distribution}
  \vspace{-0.25cm}
\end{figure*}

\subsection{Empirical Study}
\label{Human Baseline}
\begin{figure}[h]
  \vspace{-2pt}
  \includegraphics[width=\linewidth]{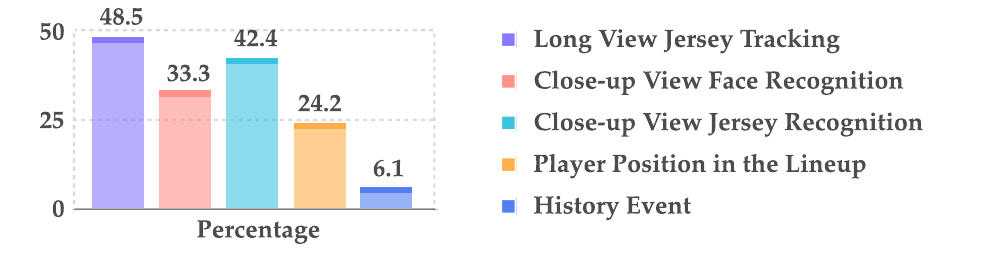} 
  \caption{Human reliance for the ``Player Guessing'' game.}
  \label{fig:human baseline}
\end{figure}

Given a soccer video segment and a piece of anonymized commentary, we develop a ``Player Guessing'' game and recruit three expert soccer commentators to take part in the study on five matches.
During the game, once the participants get a correct answer, they are request to recall the clues they rely on to pinpoint the player. 
Consequently, human participants have an accuracy of \textbf{96.3\%}. As illustrated in Fig.~\ref{fig:human baseline}, the correct identifications are made through various cues. Notably, \textbf{84.5\%} of the correct identifications were done by more than just long view player tracking, indicating the necessity of involving fine-grained visual and contextual information in player alignment. 


\subsection{Internal Game Context Construction}
\label{Context Game State Reconstruction}
The context game state $C_{GS}$ iteratively updates throughout the game. 
It mainly includes the temporal team line-up $\mathcal{T}$, $KeyEvent$ timeline, and $HistoryEvent$ timeline. 
For a match between the home team $Team_h$ ($Color_h$ jersey) and away team $Team_a$ ($Color_a$ jersey), the construction of each component is elaborated in 
Appendix~ Sec.\ref{Implementation Details in Stage I}-A.

\subsection{Fine-grained Shot Analysis} 
\label{Fine-grained Visual Details Extraction}
As shown in Fig.~\ref{fig:view shots distribution},
shot views in a soccer video are usually categorized into four classes: long view, medium view, close-up view, and out-of-field view~\cite{rafiq2020scene}.
In the soccer game sportscasting footage, 
players appearing in the preceding or following close-up and medium views are more likely to be related to events recently happened on the field.
Therefore, they provide potentially identity information to pinpoint the anonymized player. 
For the purpose of fine-grained details extraction, ``medium view'' and ``close-up view'' are collectively referred to as ``close view'' in the following sections. 
We apply several processing steps to capture the fine-grained visual details $Scene$ of the input video, including (1) Shot boundary detection with view classification, (2) Player face recognition, (3) Player jersey recognition and (4) Team affiliation detection. 
See Appendix Sec.~\ref{Details of the Fine-grained Shot Analysis} for details in each procedure.

\subsection{Frame-Level Event Grounding Guidance}
\label{Frame-Level Event Grounding Guidance}
A 30-second video segment typically contains far more events than the described one, as soccer video clips are not trimmed at the event level but centered around the ground-truth timestamp. 
For example, a corner kick may seamlessly shift into open play.
Therefore, identifying the correct event within this fixed-length segment is essential.

However, foundational video-LMMs often struggle to capture complex multi-player dynamics in the soccer domain.
To address this, we leverage the state-of-the-art soccer commentary model, MatchVision~\cite{rao_towards_2024}, to obtain prior knowledge between videos and commentary events.
Since the precise commentary response can only be generalized from the accurately grounded video segments, we obtain a frame-wise vector $\mathcal{W}$ from the Q-former's cross-attention layers of MatchVision to represent the relevance of each video frame to the generated narration 
(Details provided in Appendix Sec.~\ref{Details of the Frame-Level Event Grounding Guidance}).
It can be seen as the arousal of each frame in generating the commentary text, which serves as a frame-level event grounding guidance.

\subsection{Instruction Tuning with video-LMM}
\label{Answer-Guided Reasoning}
Given the video $\mathcal{S}$ and anonymized commentary $\mathcal{C}_A$, with $ Context = \{ C_{GS}(t),Scene(t),\mathcal{W}(t)\}$ as prompt, 
we leverage Video-LMM's visual understanding and reasoning capability to infer the correct answer $Player_g$ and his team affiliation $Team_{g}$ (example shown 
in Appendix~Fig.~\ref{fig:visual reasoning}).
Due to the suboptimal performance of the base video-LMM model, 
we employs two kinds of fine-tuning strategies to fine-tune the base model. Firstly, we adopt SFT to establish task-specific grounding. 
We further implement the GRPO approach to further enhance the model's compositional reasoning capability and reward alignment in this single choice problem.

To prepare the training data for this visual reasoning task, we use the answer-guided reasoning strategy to decompose the problem with \textit{Chain-of-Thought} reasoning. 
Through this process, we generate a train set guided by the correct answer reasoning procedure to bootstrap reasoning with reasoning~\cite{zelikman_star_2022}.
For data augmentation, we query about the [Team] and [Player] separately with different CoT strategies
(See prompt details in Appendix Sec.~\ref{Implementation Details in Stage I}-J).


%% file: paper/content/4-knowledge-enhanced.tex

\subsection{Knowledge Application in Sportscasting}
\label{Knowledge Reference in Broadcast Commentary}

\begin{figure}[h]
\vspace{-4mm}
  \includegraphics[width=\columnwidth]{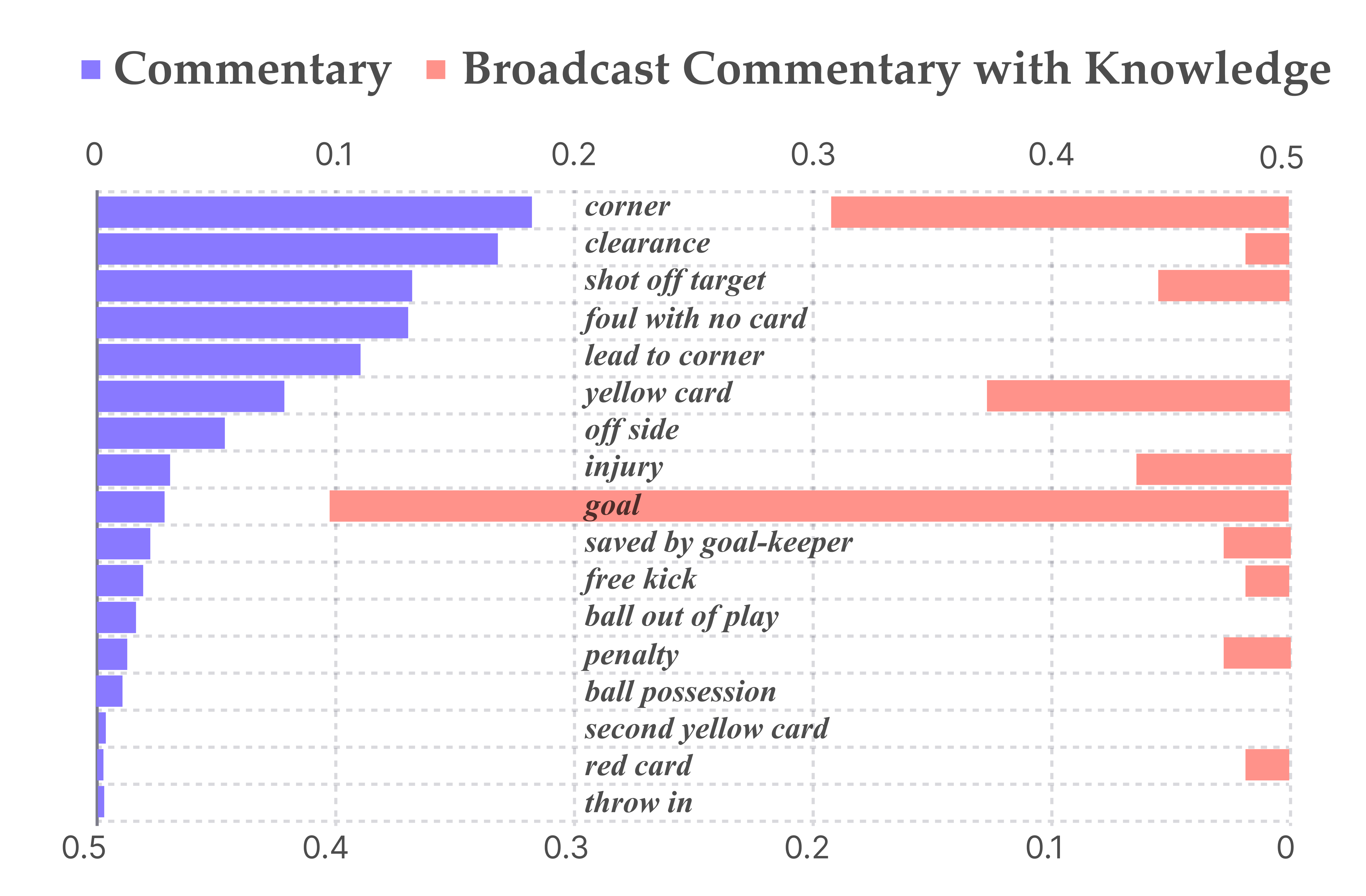}
  \caption{Distribution of commentary at event level.}
  \label{fig:commentary distribution}
\end{figure}

Live televised commentary exhibits diverse styles due to factors like the commentators' standpoint, oral habits, personality, and cultural background. 
While it is hard to reconcile these differences, human commentators share a mutual preference for incorporating contextual knowledge to offer explanations and comments on the visual scenes.

We compared the knowledge injection in live text commentary and audio transcriptions of live text commentary~\cite{li_multi-modal_nodate}. 
Inspired by FLARE~\cite{jiang_active_2023}, knowledge is extracted by LLM through paired explicit query formulations. We prompt GPT-4o to identify the expert knowledge embedded within the commentary. Additionally, we employ a self-asking mechanism~\cite{press2022measuring} to generate constrained questions grounded in the extracted knowledge.
It is found that 15.02\% of the live televised commentary contains various knowledge, whereas only 3.16\% of the live text commentary includes knowledge and is mostly limited to game score updates.
The distribution 
is illustrated in Fig.~\ref{fig:commentary distribution}. The major events where human commentators tend to reference knowledge include $Goal$, $Corner$, and $Card$.
Furthermore, based on the source of the information,  knowledge are categorized into \textbf{external knowledge}, which rely on the historical statistics from other games, and \textbf{internal knowledge}, which refers to the static background information and dynamic updated events from the current game.


\subsection{External Soccer KAG System Modification}
\label{External Soccer RAG System Modification}

The soccer KAG system, built upon SoccerRAG~\cite{strand_soccerrag_2024}, is designed for acquiring external statistics.
To better emulate human commentator behavior, we propose several key enhancements.
Firstly, we expanded the set of query exemplars in high-frequency commentary-related questions to enhance accuracy in the transition between natural language and SQL queries.
We also developed augmented data schema to add support in finer-grained event-level queries, such as own-goal and penalty for players and teams.
Furthermore, we designed strict temporal constraints within the query generation schema to ensure future matches excluded from retrieval. It maintains the temporal integrity of the database and prevents data leakage during inference.

\subsection{Internal Game Context Tracking}
\label{Internal Game Context Knowledge Construction}

The internal game context module tracks in-game information essential for commentary generation. Its construction follows an approach similar to Sec.~\ref{Context Game State Reconstruction}, with adjustments guided by the human commentators' reference habits~(Sec.~\ref{Knowledge Reference in Broadcast Commentary}).
Detailed background about the mentioned player, such as nationality and height, are provided in this module.
Fine-grained information such as the goal scorer, assisting players, and the specific method of goal achievement (e.g., penalty, header, or own goal) are also captured.

\subsection{Commentary Refinement with LLM}
\label{Commentary Refinement with LLM}

With the stage I commentary $\mathcal{C}_{EA}$ describing the basic current event, we use GPT-4o to generate questions related to external statistics. The responses from the KAG system are then double checked with LLM to discard the invalid answers with repetition or incorrect temporal range. 
In addition, internal game context knowledge, such as the key timeline of the mentioned event and detailed player information, are explicitly prompted for the model to quote.
This process enhances the accuracy and insight of the commentary, making it more aligned with live televised commentary.

%% file: paper/content/5-experiments.tex
In this section, we first introduce the experimental setup in Sec.~\ref{Experimental Setup}, then unfold two series of experiments for each stage of \textsc{GameSight}.
As to stage I, Sec.~\ref{Accuracy of Entity Alignment} shows experiments on the accuracy of entity alignment. 
For stage II, 
we evaluate the performance of \textsc{GameSight} refined commentary in Sec.~\ref{Commentary Refinement Evaluation} from segment-level: baseline comparison, knowledge accuracy, and game-level: structural similarity, sentiment polarity and contextual relevance.

\subsection{Experimental Setup}
\label{Experimental Setup}
\textbf{Stage I.} 
We use the test set from SN-Caption-test-align (49 games, 2,305 segments) and train set from MatchTime (419 games, 18,826 segments).
Zero-shot experiments on video-LMMs are conducted with video fps=1 and video quality in 720p (1,280$\times$720 pixels) for the clearest visual details.
Instruction-tuning experiments are conducted on 4 $\times$ H100 (80G) GPUs, with fps=1 and video quality in 180p (320$\times$180 pixels) due to the resource limit. We use Qwen2.5-VL-7B-Instruct as the backbone model.
SFT experiments are trained for 2 epochs with a learning rate of 1e-4. LoRA~\cite{hu2021loralowrankadaptationlarge} is adopted with a rank of 8. GRPO experiments are trained each for 600 steps with binary reward. Learning rate is 1e-6 and $\beta$ coefficient is 0.1. Following ~\cite{wang-2025-open-r1-video}, we only utilize video, query, and the ground truth answer for training. 

We report $Player$ alignment accuracy, and derive $Team$ alignment accuracy based on the predicted player’s team.
$Player_c$ and $Team_c$ denote settings where the correct team affiliation is given, reducing the candidate search space.

\noindent\textbf{Stage II.}
As to segment-level commentary generation, we use SN-Caption-test-align (live text) and Goal benchmark~\cite{qi_goal_2023} (live televised) as two versions of ground truth. The state-of-the-art soccer commentary models MatchVoice~\cite{rao_matchtime_2024} and MatchVision~\cite{rao_towards_2024} are chosen as the baselines.
We employ BLEU~\cite{papineni2002bleu}, CIDER~\cite{vedantam2015cider}, METEOR~\cite{banerjee2005meteor} and ROUGE~\cite{lin2004rouge}
as evaluation metrics. All commentary are anonymized for a fair comparison.
For the internal knowledge accuracy evaluation, we conduct knowledge refinement using ICL knowledge only, and extract the referred internal statistics using the same prompt as Sec.~\ref{Knowledge Reference in Broadcast Commentary}. Catering to external knowledge accuracy, we manually tailor a set of 500 statistical questions reflecting human commentators' reference habits. Detailed samples are 
elaborated in Appendix~ Sec.\ref{Implementation Details in Experiments}-A.2.

For game-level evaluation, we consider two baselines: the game-wise concatenated live text commentary~\cite{deliege_soccernet-v2_2021}, and the whole 90 minutes sportscast's ASR transcription~\cite{gautam_soccernet-echoes_2024}. 
In quantitative analysis, we adopt human evaluation, sentiment polarity, and the \textit{Coh-Metrix}~\cite{mcnamara2014automated} in discourse analysis to evaluate the coherence and contextual relevance.
For the structural composition of commentary, sport communication experts~\cite{zhang2017shanghai} defines three logical categories: description, explanation, and comment, with the criteria that description in live televised commentary should not exceed 50\%. We use GPT-3.5-turbo to conduct the classification.

\subsection{Accuracy of Entity Alignment}
\label{Accuracy of Entity Alignment}

\noindent\textbf{Exp.~1: Zero-shot accuracy test of video-LMMs and LLMs.}
We select Qwen2.5-VL-7B-Instruct~\cite{qwen2.5-VL}, InternVL3~\cite{zhu2025internvl3exploringadvancedtraining},  VideoLLaMA3-7B~\cite{zhang2025videollama3frontiermultimodal}, and LLaVA-OneVision~\cite{li2024llavaonevisioneasyvisualtask} as the representative of open-sourced video-LMMs, Gemini 2.5-pro~\cite{google2023gemini} for close-sourced video-LMM, GPT-4o~\cite{openai2024gpt4o}, GPT-o1-preview~\cite{openai2024introducing} and DeepSeek-R1~\cite{deepseekai2025deepseekr1incentivizingreasoningcapability} for close-sourced LLMs. We additionally use the top-1/3 accuracy for player alignment.
Tab.~\ref{table_zero_shot} demonstrates the zero-shot abilities of all kinds of video-LMMs and LLMs in the player alignment reasoning task.
Although the contextual and visual clues can be organized in text format, LLMs performance falls short of the open-sourced video-LMMs, indicating the importance of involving visual modality to further observe the related scene.
While Qwen2.5-VL lags behind close-sourced video-LMM in performance, it outperforms open-sourced models and most LLMs, which serves as the superior backbone for further fine-tuning.

\begin{table}[ht]
\centering
\caption{Zero-shot performance of video-LMMs and LLMs. Bolded and \underline{underlined} for the 1st and 2nd largest value.}
\resizebox{\columnwidth}{!}{
\begin{tabular}{lcccccc}
\toprule
\textbf{Model} & \textbf{$Player@1$} & \textbf{$@3$} & \textbf{$Team$} & \textbf{$Player@1_{c}$} & \textbf{$@3_{c}$} & \textbf{$Team_{c}$}\\
\midrule
Qwen2.5-VL & \underline{35.4} & \underline{39.3} & 70.9 & \underline{43.8} & 48.6 & 91.4 \\
InternVL3 & 26.3 & 39.1 & \underline{72.2} & 30.8 & 47.4 & 92.0 \\
VideoLLaMA3 & 17.2 & 27.5 & 63.7 & 19.4 & 36.4 & 83.7 \\
LLaVA-OV & 18.7 & 18.9 & 66.7 & 21.7 & 21.9 & 87.9 \\
\hdashline
Gemini 2.5-pro & \textbf{52.6} & \textbf{55.9} & \textbf{71.2} & \textbf{61.3} & 62.4 & 96.9 \\
\hdashline
GPT-4o & 21.2 & 24.0 & 49.5 & 37.7 & 45.0 & \underline{98.0} \\
GPT-o1 & 30.7 & 37.9 & 53.9 & 40.8 & \underline{62.8} & \textbf{98.2} \\
DeepSeek-R1 & 23.8 & 28.6 & 42.9 & 35.9 & \textbf{63.2} & 97.6 \\
\bottomrule
\end{tabular}
}
\label{table_zero_shot}
\end{table}

\noindent\textbf{Exp.~2: Instruction tuning with video-LMM.}
Tab.~\ref{table_sft} validates the effectiveness of the proposed fine-tuning strategy.
Introducing SFT markedly improves $Player$ accuracy by 27.3\% compared with base model, demonstrating strong task grounding with SFT.
GRPO alone is less effective than SFT in boosting the performance, but it still shows that structured reasoning benefits from reward-driven alignment with a modest improvement.
The combination of both SFT and GRPO leads to the best overall performance, with substantial gains in all categories besides $Team_c$. 
The synergy between SFT and GRPO proves crucial for addressing the multifaceted challenges of visual reasoning in video-LMMs.

\begin{table}[ht]
\centering
\caption{Experiment results of fine-tuning on Qwen2.5-VL.}
\label{table_sft}
\resizebox{\columnwidth}{!}{
\begin{tabular}{lcccc}
\toprule
\textbf{Training Strategy} & 
\textbf{$Player$}~$\uparrow$ & \textbf{$Team$}~$\uparrow$ & \textbf{$Player_{c}$}~$\uparrow$ & \textbf{$Team_{c}$}~$\uparrow$
\\
\midrule
Qwen2.5-VL \#1 & 35.9 & 69.6 & 43.9 & 92.5 \\
\hdashline
\#1 + SFT & \underline{63.2} & \underline{81.7} & \underline{67.0} & \underline{91.6} \\
\#1 + GRPO & 43.7 & 72.9 & 50.7 & \textbf{92.7}  \\
\textbf{\#1 + SFT+GRPO~(Ours)} & \textbf{71.1} & \textbf{84.7} & \textbf{75.9} & 88.7 \\
\bottomrule
\end{tabular}
}
\end{table}

\begin{table}[ht]
\centering
\caption{Comparison with baselines.}
\label{table_captioning_eval}
\resizebox{\columnwidth}{!}{
\begin{tabular}{lcccc}
\toprule
\textbf{Model} & \textbf{BLEU} & \textbf{METEOR} & \textbf{ROUGE} & \textbf{CIDEr} \\
\midrule
\multicolumn{5}{c}{\textit{Live text as GT}} \\
\midrule
MatchVoice       & 15.182     & 22.142     & 18.262     & \underline{13.514}     \\
MatchVision      & \textbf{28.311} & \underline{25.827}     & \underline{25.173}     & \textbf{27.143} \\
\textsc{GameSight}~\textbf{(ours)} & \underline{20.320}     & \textbf{27.160} & \textbf{27.968} & 0.887      \\
Live televised   & 4.038      & 11.778     & 6.272      & 0.001      \\
\midrule
\multicolumn{5}{c}{\textit{Live televised as GT}} \\
\midrule
MatchVoice       & \underline{8.615}      & \textbf{10.067} & 8.247      & 0.113     \\
MatchVision      & 4.214      & 7.945      & \underline{8.946}      & \underline{0.369}     \\
\textsc{GameSight}~\textbf{(ours)} & \textbf{17.409} & \underline{9.051}      & \textbf{10.497} & \textbf{3.442} \\
\bottomrule
\end{tabular}
}
\vspace{-4mm}
\end{table}

\subsection{Commentary Refinement Evaluation}
\label{Commentary Refinement Evaluation}

\noindent\textbf{Exp.~3: Segment-level commentary baseline comparison.}
Tab.~\ref{table_captioning_eval} shows that \textsc{GameSight} has competitive performance across both settings. It has relatively lower similarity to live-text commentary compared to MatchVision in BLEU and CIDER, since the knowledgeable commentary has a naturally lower similarity to the original live-text commentary, as shown in the last row under \textit{Live-text as GT}. They are usually longer and more analytical, differs significantly in natural language style. While, \textsc{GameSight} outperforms others when using live televised as GT, showing a better alignment to the human-centric TV commentary.

\noindent\textbf{Exp.~4: Segment-level accuracy of knowledge reference.}
As shown in Fig.~\ref{table_knowledge_reference}, SoccerKAG improves external knowledge accuracy by 16.11\% compared to SoccerRAG, which primarily attributes to the enhanced query exemplars and time constrains in query schema. Regarding the internal game context, we further break down the test into two categories: goal-related information~($ICL_{goal}$) and other context aspects~($ICL_{other}$).
\textsc{GameSight} presents high accuracy
particularly in goal events, as the goal information is specifically quantified by the score. A slight drop in accuracy is observed when referencing other internal context elements, for instance, fouls and corner kicks, as the model independently accounts for various demanded statistics.

\begin{table}[ht]
\centering
\caption{Results on knowledge reference accuracy.}
\label{table_knowledge_reference}
\resizebox{\columnwidth}{!}{
\begin{tabular}{ccccc}
\toprule
 & $SoccerRAG$ & $SoccerKAG$ & $ICL_{goal}$ & $ICL_{other}$ \\
\midrule
\textit{Acc.}~$\uparrow$ & 64.60 & \textbf{81.80}  & \textbf{98.76} & \textbf{90.74}\\
\bottomrule
\end{tabular}
}
\end{table}

\noindent\textbf{Exp.~5: Game-level quantitative results in commentary analysis.}
\textit{Coh-Metrix}~\cite{mcnamara2014automated} is a discourse analyze tool that 
includes various indicators to evaluate discourse coherence.
With the focus on contextual relevance, we adopted two key indicators, deep coherence and anaphor overlap.
\textit{Deep coherence} reflects the degree of text containing intentional connectors when causal and logical relationships present. As shown in Tab.~\ref{table_coh_metrix}, televised and \textsc{GameSight} commentary have higher deep coherence that helps readers form a more engaging and in-depth understanding of causal events, processes, and behaviors during the game.
\textit{Anaphor overlap} measures the overlap between nouns and pronouns in adjacent sentences, indicating the semantic continuity within the passage by pointing back to the context. 
Televised and \textsc{GameSight} commentary show higher anaphor overlap, indicating better contextual relevance. 
\textit{Sentiment Score}~\cite{hutto2014vader} is a polarity score reveals the sentiment polarity (1=positive, -1=negative). Live text has almost neutral sentiment, while televised and \textsc{GameSight} commentary provide positive and vivid atmosphere.
The MOS test also shows that audience prefers our commentary than the original live text version.

\begin{table}[ht]
\centering
\caption{Results in commentary analysis. DC, AO, SS refer to \textit{Deep Cohesion}, \textit{Anaphor Overlap}, and \textit{Sentiment Score}.}
\label{table_coh_metrix}
\resizebox{0.9\columnwidth}{!}{
\begin{tabular}{lcccc}
\toprule
\textbf{Method} & \textbf{DC}$\uparrow$ & \textbf{AO}$\uparrow$ & \textbf{SS}$\uparrow$ & \textbf{MOS}$\uparrow$\\
\midrule
MatchVoice & 0.556 & 0.164 & 0.159 & 2.27\\
MatchVision & 0.631 & 0.156 & 0.066 & 2.72 \\
Live Text & 0.474 & 0.174 & -0.059 & 3.14 \\
\textsc{GameSight}~\textbf{(ours)} & \textbf{0.842} & \underline{0.213} & \underline{0.194} & \underline{4.08} \\
Live Televised & \underline{0.813} & \textbf{0.234} & \textbf{0.254} & \textbf{4.22} \\
\bottomrule
\end{tabular}
}
\end{table}

\begin{figure}[h]
  \includegraphics[width=\linewidth]{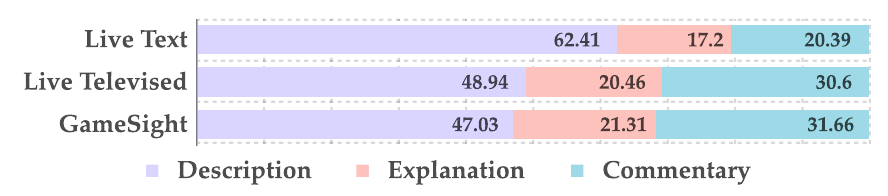}
  \caption{Commentary structural composition in percentage.}
  \label{fig:structural composition}
\end{figure}

\noindent\textbf{Exp.~6: Game-level commentary structural similarity.}
As shown in Fig.~\ref{fig:structural composition}, while live text commentary remains small proportion of explanation and commentary, \textsc{GameSight} has a similar structure with live televised commentary, which meets the criteria from ~\cite{zhang2017shanghai} with less than 50\% description, indicating the depth and insights in the commentary with appropriate logical composition.
\vspace{-4mm}
\vspace{-4mm}



%% file: paper/content/6-conclusion.tex
In this work, we proposed \textsc{GameSight} to tackle soccer commentary generation as a knowledge-enhanced visual reasoning task.
It enables automatic commentary generation with precise entity and contextual knowledge, which leans towards the real-world live televised commentary that provides the audience with informative and engaging experience.
We believe our work paved way for leveraging video-LMMs in fine-grained sports analysis.

%% file: paper/content/7-appendix.tex
\begin{figure*}[h]
  \includegraphics[width=\linewidth]{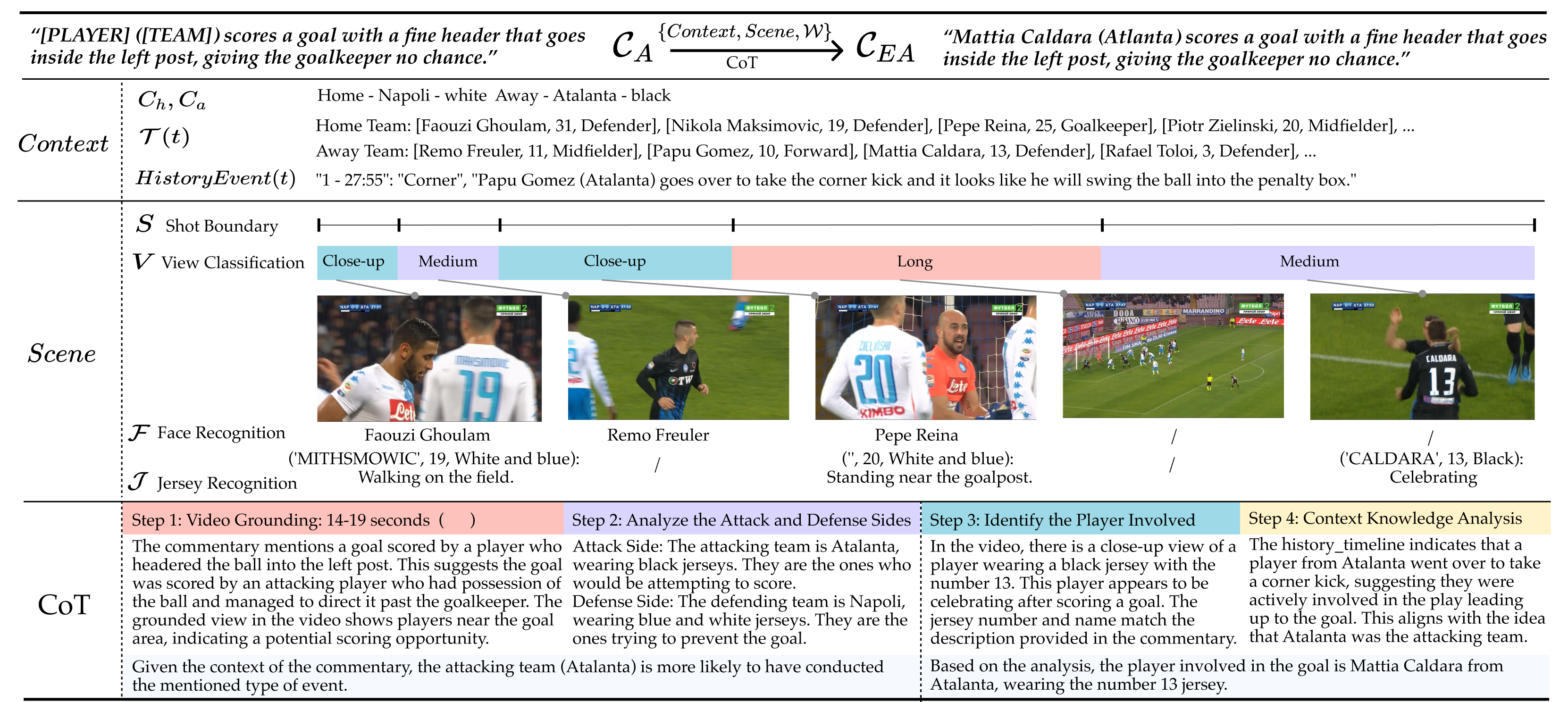}
  \caption{An example of CoT procedure in stage I: visual reasoning for entity alignment with \textsc{GameSight}.}
  \label{fig:visual reasoning}
\end{figure*}

\section{Limitations and Future Works}

While \textsc{GameSight} marks a significant advancement in soccer commentary generation by addressing both visual reasoning and knowledge enhancement, there are several limitations and potential avenues for future improvement.

\subsection{Generalization Across Sports:} \textsc{GameSight} has been designed specifically for soccer commentary generation, leveraging domain-specific knowledge and event detection techniques tailored to soccer. Expanding the model to accommodate other sports with differing event structures, player roles, and contextual information will require further adaptation. Future work could explore the model's generalizability to other sports, incorporating sport-specific visual reasoning and knowledge bases.

\subsection{Real-Time Processing:} 
Real-time soccer commentary generation remains a challenge. 
Although our stage I (visual reasoning via video-LMM) is streaming-oriented, and stage II (knowledge enhancement by KAG and internal context updates) can be adapted for streaming with some potentially engineering optimizations. However, current soccer commentary generation models, the upstream model such as MatchVision  and SoccerComment, still operates on fixed-length video segments, which results in latency during live events. Future works could investigate ways to optimize the model for streaming performance while maintaining the quality of insights and relevance in commentary.

\subsection{Diverse Knowledge Sources:} While \textsc{GameSight} integrates external statistics and internal game state updates, the scope of knowledge is restricted to statistics in the database. Future research could focus on incorporating more diverse external knowledge sources, such as player health status, or social media trends, to improve the contextual awareness of the commentary by retrieval from the open web.

\subsection{Human-Like Intuition and Creativity:} \textsc{GameSight} still lacks the creative and nuanced touch characteristic of human commentators. The ability to deliver insightful, emotionally engaging, and creative commentary remains an open challenge. Incorporating deeper understanding of human-like intuition, humor, and emotional expressions into the AI's generation process could be a promising direction for future work.

\section{Data Implementation Details}
The MatchTime dataset is curated from the SoccerNet-Caption, containing 422 automatically aligned matches in train and valid set, and 49 manually curated matched, called the SN-Caption-test-align benchmark.

\subsection{Data Validation.} 

As the accuracy of the score is crucial throughout the game, we took extra steps to ensure data quality. 
The 471 matches from the original SoccerNet-v2 Dataset (initially build up from \url{www.flashscore.info}) had 133 matches with goal event omission in both labels and commentary. To address this, we crawled the goal timeline from another sports statistics websites~(\url{www.worldfootball.net}), and fixed the mismatch with 188 added missing goals.
Additionally, for the effectiveness of game state reconstruction,
we ensured that players mentioned in each commentary segment were present in the temporal line-ups.
While commentary may occasionally mention players on the bench in special cases, such as a referee's warning, we filter these to narrow the player range to the official line-up for consistency.
We removed 386 inaccessible entries due to the previous missing substitution events in SoccerNet-v2. 

\subsection{Data Filtering.}

In the end-to-end soccer commentary generation task, four types of anonymized entities are considered: [PLAYER], [TEAM], [COACH], and [REFEREE]. 
Since there are only one referee and two coaches on the court, and the [COACH] reference always come up with [PLAYER] from the same team, they are relatively convenient to detect. 
Among these, [PLAYER] and [TEAM] are the two most frequent terms in the dataset's word frequency analysis~\cite{rao_towards_2024}. 
Consequently, we prioritize the alignment of [PLAYER] and [TEAM] labels in the anonymized commentary text. 
Furthermore, to better formulate the problem, we filtered the commentary pieces with multiple entities and only preserved single-player anonymized commentary.
We obtained 18,826 commentary segments
 in train set and 2,305 pieces of commentary in test set.

\section{Implementation Details in Stage I}
\label{Implementation Details in Stage I}
\subsection{Details of the internal Game Context Construction~(Sec.~IV-B)} 

$\mathcal{T}(t)$ records the identification of twenty-two players currently active on the field, with detailed player information:
\begin{equation}
\begin{aligned}
&Player = (Name, Number, Position). \\
&Team_i = \left[Player_i^j, Coach_i \mid  j \in \left[1, 11\right] \right], i\in\{a,h\}, \\
&\mathcal{T}(t) = Team_h(t) \cup Team_a(t).
\end{aligned}
\end{equation}
Define $E(t)$ as a sequence of timestamped match events. $KeyEvent$ timeline records the timestamps and details of goals and cards events up to the current moment of the game. 
$HistoryEvent$ timeline consists of the latest $k$ pieces of commentary, reflecting the recent happenings on the field:
\begin{equation}
\begin{aligned}
&KeyEvent(t) = \left[ E(i) \mid E(i) \in \{Goal, Card\}, i < t \right],\\
&HistoryEvent(t) = \left[ E(i) \mid i \in \left[t-k-1, t\right) \right].
\end{aligned}
\end{equation}
Throughout the game, the evolving score and dynamic team line-ups are key elements that define the flow of the match.
Game score is updated when a goal occurs. Let $P$ be the scoring player:
\begin{equation}
\begin{aligned}
&Score_a \leftarrow Score_a + 1, \text{ if } E(t) \in {Goal} \text{, } P \in Team_a(t)\\
&Score_h \leftarrow Score_h + 1, \text{ if } E(t) \in {OwnGoal} \text{, }  P \in Team_a(t)
\end{aligned}
\end{equation}
Team line-up updates with each substitution, which are handled by replacing an active player $P$ with a substitute $Q$:
\begin{equation}
\begin{aligned}
&Team_i(t+1) = \left( Team_i(t) \setminus \{P\} \right) \cup \{Q\},  \\
&\text{if } E(t) \in {Substitution}, \text{ } P,Q \in Team_i(t), \text{ } i \in \{h, a\}\\
\end{aligned}
\end{equation}

\subsection{Details of the Fine-grained Shot Analysis~(Sec.~IV-C)} 
\label{Details of the Fine-grained Shot Analysis}

\noindent\textbf{(1) Shot boundary detection with view classification.}
We first split the 30-second video into a sequence of shots, denoted as $\mathcal{S} = \{S_1, ..., S_n\}$ and classified each shot into one of four shot categories with the vision language model~(VLM).
The shot categories are represented as 
\begin{equation}
  V(S_i) \in \{V_{\text{close}}, V_{\text{medium}}, V_{\text{long}}, V_{\text{audience}}\},  
\end{equation}


\noindent\textbf{(2) Player face recognition.}
We crawled the face images of 3,213 Soccernet-v2 involved players from the sourced sports statistics website to create an image database for face alignment in close views.
For each close view shot $S_i$ in the video segment, player face recognition ($\phi$) is performed to compare the faces~\{$f_{i}^1,...,f_{i}^m$\} appeared in the video frames with the candidates from current line-ups $\mathcal{T}(t)$.
The process is repeated across three key frames as \(S_{i,k}\)~(\(k=1,2,3\)) to mitigate issues such as blur and out-of-frame images caused by dynamic camera changes. The finally recognized faces are denoted as $\mathcal{F}_i$.
\begin{equation}
\begin{aligned}
&\mathcal{F}_i = \left\{\, p \in \mathcal{T}(t) \,\middle|\, \max_{k=1,2,3} \phi(f_{i,k}^j, p) > \tau,\ j \in [1, m] \,\right\}
\end{aligned}
\end{equation}

\noindent\textbf{(3) Player jersey recognition.}
The accuracy of VLM's optical character recognition (OCR) capability decreases as fine-grained details like names and jersey numbers are small, blurry, or overlapped.
Therefore, we only instruct the VLM to detect in close view shots $S_i$.
To avoid irrelevant or general scene descriptions, the OCR and captioning results are restricted to the following format: 
\begin{equation}
\mathcal{J}_i = \left\{(Name_P, Number_P, Color_P): Action_P\right\}.
\end{equation}
The ultimate fine-grained visual details extracted from a 30-second video segment is:
\begin{equation}
    Scene(t) = \{V(S_i) \cup \mathcal{F}_i \cup \mathcal{J}_i \mid i \in \left[1, n\right]\}
\end{equation}
\noindent\textbf{(4) Team affiliation detection.}

Team affiliation detection refers to link a team to the jersey color worn during the match. 
To accurately determine team affiliation, we leverage the knowledge of traditional team colors for VLM to cross-check within the video. 
LLM is further employed to merge similar description in color terms (e.g., ``red and blue'' with ``blue and red striped'') and filter out corner cases. 
When ambiguities arise (e.g. distinguishing the ``blue'' team between a ``light blue'' and ``dark blue'' match), we identify the team by its unique jersey number spotted in video, which belongs exclusively to one side of the current line-ups.
Finally, we manually verify the remaining instances to determine the home team color $C_h$ and away team color $C_a$.

\noindent\textbf{(5) Implementation Details.}

We use ContentDetector to detect the scene boundary with threshold=16 and video fps=25.
Player face recognition is conducted with the face recognition package\footnote{\url{https://github.com/ageitgey/face_recognition}.} with threshold $\tau$=0.6. 
Hyperparameter k in the $HistoryEvent$ construction is set to be 1. An example of stage I visual reasoning is shown in Fig.~\ref{fig:visual reasoning}.

\subsection{Details of the Frame-Level Event Grounding Guidance~(Sec.~IV-D)}

\label{Details of the Frame-Level Event Grounding Guidance}
A 30-second video segment typically contains far more events than the described one, as soccer video clips are not trimmed at the event level but centered around the ground-truth timestamp. 
For example, a corner kick may seamlessly shift into open play.
Therefore, identifying the correct event within this fixed-length segment is essential.

However, foundational video-LMMs often struggle to capture complex multi-player dynamics in the soccer domain.
To address this, we leverage the state-of-the-art soccer commentary model, MatchVision~\cite{rao_towards_2024}, to obtain prior knowledge between videos and commentary events.
Since the correct commentary response can only be generalized from the accurately grounded video segments, we extract its Q-former’s cross-attention layers, which query frame-wise features through temporal self-attention, generating prefix embeddings for the LLM decoder.

We first compute the average cross attention between queries and video frames. Let \( A^{(\ell, h)}_{q,n} \) be the attention weight for query \( q \) and frame \( n \) at layer \( \ell \) and head \( h \). Averaging over all \( L \) layers and \( H \) heads, we obtain:
\[
\bar{A}_{q,n} = \sum\nolimits_{\ell=1}^{L} \sum\nolimits_{h=1}^{H} A^{(\ell, h)}_{q,n} \big/ (L \cdot H)
\]

We then compute the query importance using the L2 norm of the query output \( v_q \), where \( Q \) is the number of queries:
\[
\alpha_q = \|v_q\|_2 \big/ \sum\nolimits_{q'=1}^{Q} \|v_{q'}\|_2
\]

The final aggregated frame-wise attention is given by:
\[
\mathcal{W} = \{a_n\}_{n=1}^{N}, \quad a_n = \sum\nolimits_{q=1}^{Q} \alpha_q\, \bar{A}_{q,n}
\]
Consequently, we obtain a frame-wise vector $\mathcal{W}$ representing the relevance of each video frame to the generated narration.
It can be seen as the arousal of each frame in generating the commentary text, which serves as a frame-level event grounding guidance.
\\
\subsection{Validation of Shot Analysis} 

\begin{figure}[h]
  \includegraphics[width=\linewidth]{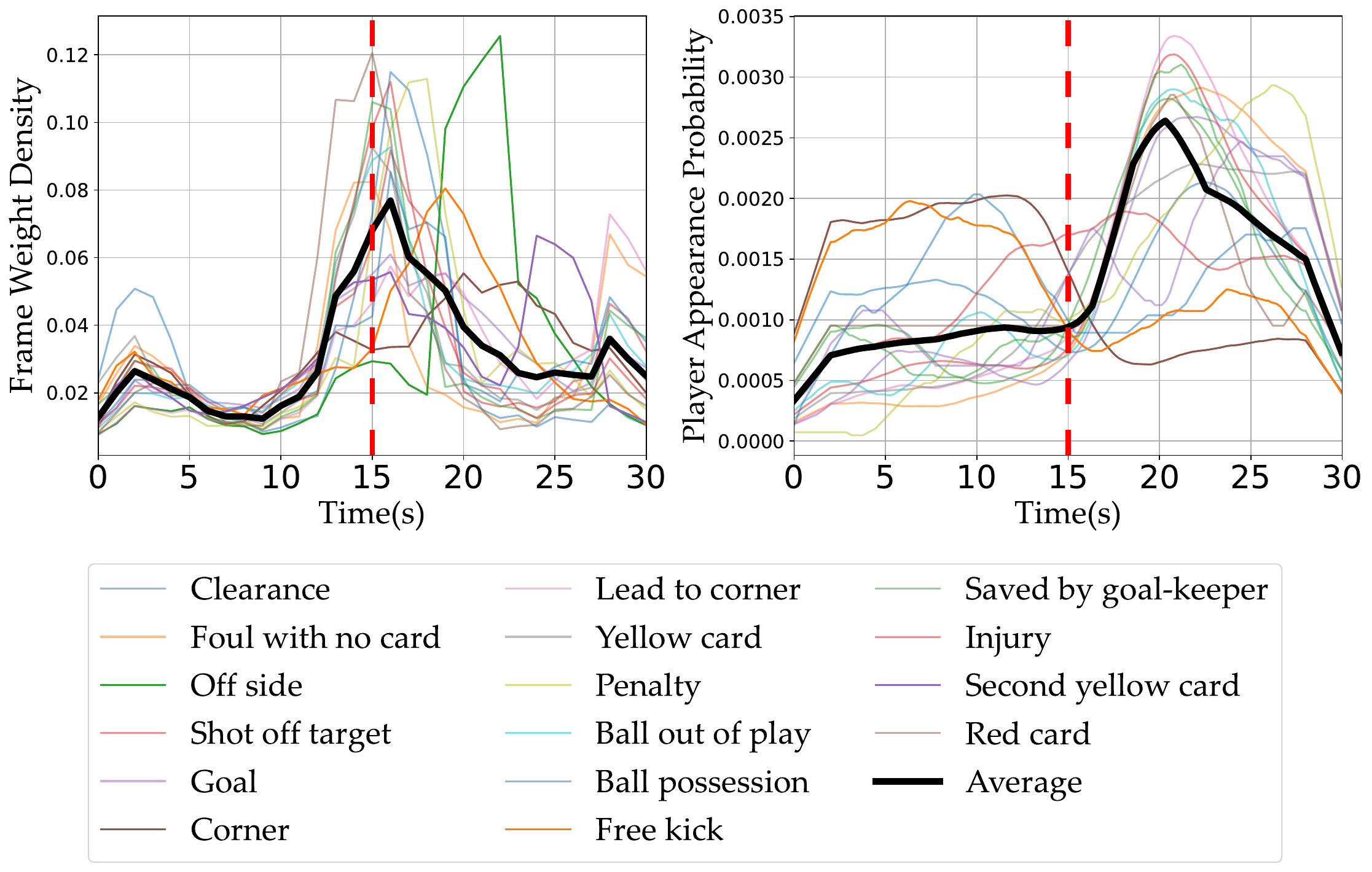}
  \caption{Test set frame-level distribution of \textit{Left:} grounded video segments and \textit{Right:} close views with valid clues.}
  \label{fig:frame_scene}
\end{figure}

\begin{figure}[h]
  \includegraphics[width=\linewidth]{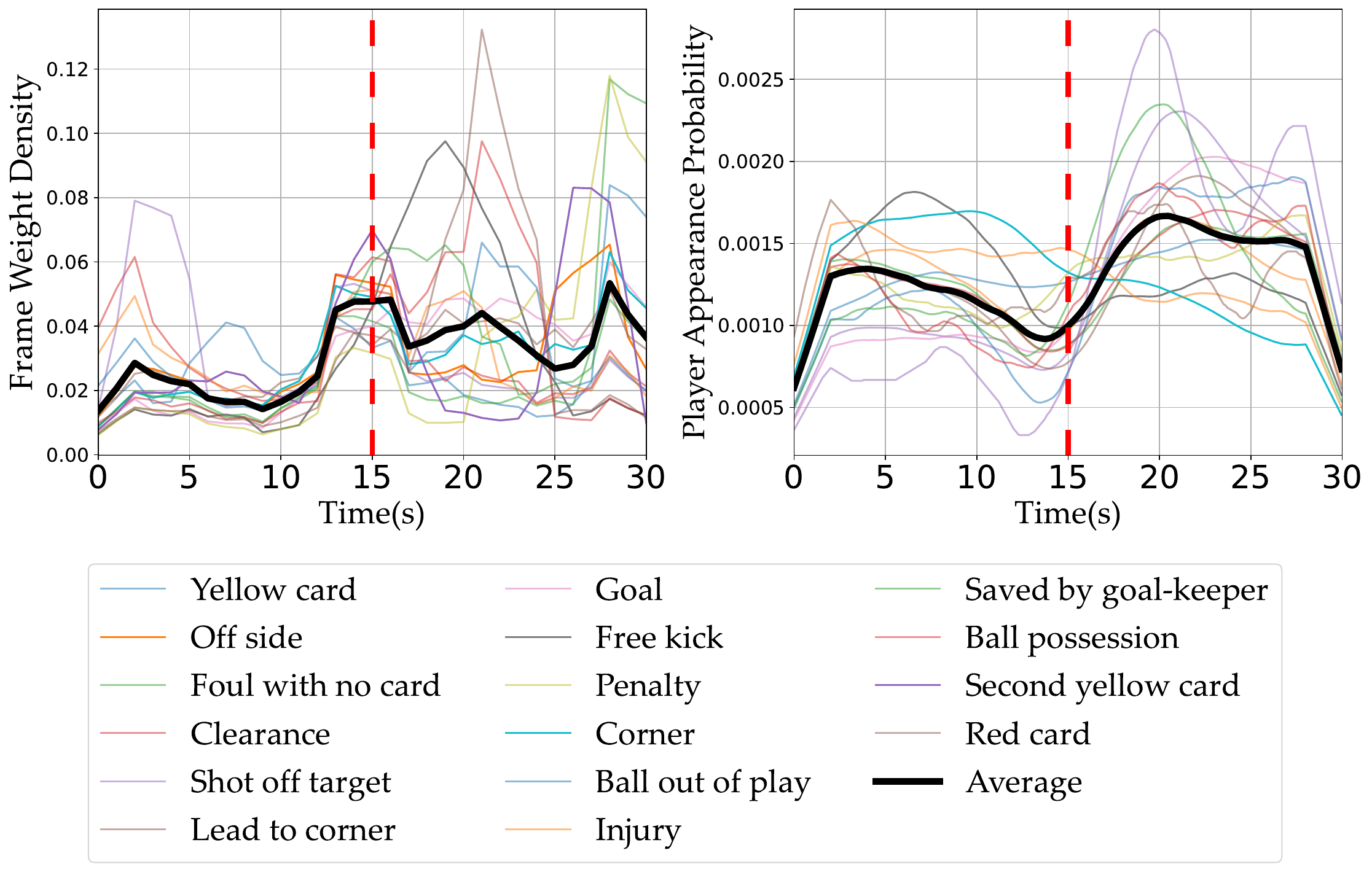}
  \caption{Train set frame-level distribution of \textit{Left:} grounded video segments and \textit{Right:} valid close-up shots.}
  \label{fig:frame_scene_trainset}
\end{figure}

We start by ensuring the effectiveness of visual details extraction (Sec.~IV-C) to provide valid clues. We calculate the distribution of the top five frames in $\mathcal{W}(t)$ as the guidance of event video grounding in the 30 seconds videos (fps = 1 for $\mathcal{W}(t)$), as well as the distribution of close views where the correct player's face or jersey is recognized in the 750 frames (fps = 25 for shot boundary detection). 

Result on the test set is shown in Fig.~\ref{fig:frame_scene}. 
Given that the videos has manually checked timestamps and are composed by the 30 seconds centered with the timestamps, the event should be best grounded at around 15 second, which is aligned with the average climax of $\mathcal{W}(t)$ distribution, showing that the proposed $\mathcal{W}(t)$ has valid guidance of event video grounding. 
Moreover, the valid clues in close views often appear in the latter part of the video, which is also reasonable because players involved in the past event are usually given following close shots. 
Corner and free kick events are also exceptions, where close-up views for the event executors are usually given in advance, reflected by the opposite tendency of the orange and brown line. The corresponding figure for the train set is shown in Fig.~\ref{fig:frame_scene_trainset}. 

Since the timestamps of train set commentary are not manually curated but automatically aligned by the MatchTime model, the effective video segments in the training set do not exhibit the clear pattern seen in the test set, where the segments are mainly concentrated around the 15-second mark. Instead, the train set segments are more irregularly distributed across 30-second intervals, further emphasizing the necessity of leveraging the pre-trained end-to-end model to provide frame-level prior guidance of video grounding.


\subsection{Ablation study on the clue components.}
We conduct ablation test to detect whether including each component positively contributes to the visual reasoning in alignment accuracy, to decide the best composition of fine-tuning prompt. 
It can be observed from Tab.~\ref{table_ablation} that the contribution of various components to the visual reasoning task differ, while most of them are positive. 
The face information and CoT modules are especially useful with a significant drop in accuracy when excluded, highlighting the crucial role of close views analysis and logical reasoning steps in understanding complicated player states and actions. 
Furthermore, when the correct team affiliation is provided ($Player_c$ and $Team_c$), the accuracy for both tasks generally improves, suggesting that additional contextual information enhances the model's reasoning ability. 
However, the high accuracy in $Team_c$ without $CoT$ shows that deep reasoning has slightly effects the instruction following capability that the zero-shot model possesses. 
Moreover, since the exclusion of KetEvent Timeline ($-KeyE$) outperforms the baseline, suggesting that it is more likely to be a distraction than providing valid clues, this component is excluded from the contextual clues in the entity alignment experiments in Sec.~VI-B.

\begin{table}[ht]
\centering
\caption{Experiment results of the ablation test on the contextual and visual components with Qwen2.5-VL. \underline{Underlined} results are above the baseline. $C$, $\mathcal{F}$, $\mathcal{J}$, $\mathcal{W}$ refer to $Color$, ${\mathcal{F}ace}$, ${\mathcal{J}ersey}$, $\mathcal{W}eight$ respectively.}
\label{table_ablation}
\resizebox{\columnwidth}{!}{
\begin{tabular}{lcccccccc}
\toprule
 & \textbf{All} & $-C$ & ${\mathcal{F}}$ &  $-{\mathcal{J}}$ &$-\mathcal{W}$ & $-KeyE$ & $-HistoryE$ & $-$CoT \\
\midrule
$Player$ & \textbf{35.5} & 34.8 & 23.5 & 34.9 & 34.7 & \underline{35.5} & 35.3 & 26.1 \\
$Team$ & \textbf{69.5} & 68.7 & 63.8 & \underline{70.6} & 68.8 & \underline{71.0} & 69.4 & 65.6 \\
\hdashline
$Player_c$ & \textbf{43.6} & 42.2 & 30.2 & 43.4 & 42.2 & \underline{43.8} & \underline{43.7} & 38.5 \\
$Team_c$ & \textbf{91.9} & 90.7 & 89.6 & \underline{92.9} & 90.7 & 91.6 & 91.5 & \underline{95.4} \\
\bottomrule
\end{tabular}
}
\end{table}

\subsection{Fine-tuning with different version of timestamps.}

We present the SFT training results on the train set video segments clipped by enhanced MatchTime timestamps and the original SoccerNet-Caption timestamps respectively. Tab.~\ref{table_timestamps} shows that the enhanced timestamps automatically provided by MatchTime alleviated the video-text misalignment issue present in the original data.

\begin{table}[ht]
\centering
\caption{Experiment results with different version of timestamps.}
\label{table_timestamps}
\begin{tabular}{lcc}
\toprule
\textbf{Timestamp Version} & $Player$~$\uparrow$ & $Team$~$\uparrow$\\
\midrule
MatchTime & \textbf{63.2\%} & \textbf{81.7\%} \\
SoccerNet-Caption & -1.6\% & -2.6\% \\
\bottomrule
\end{tabular}
\end{table}

\subsection{Visual Interface of the ``Player Guessing'' Game (Sec.~IV-A)}

\begin{figure*}[h]
  \includegraphics[width=0.9\textwidth]{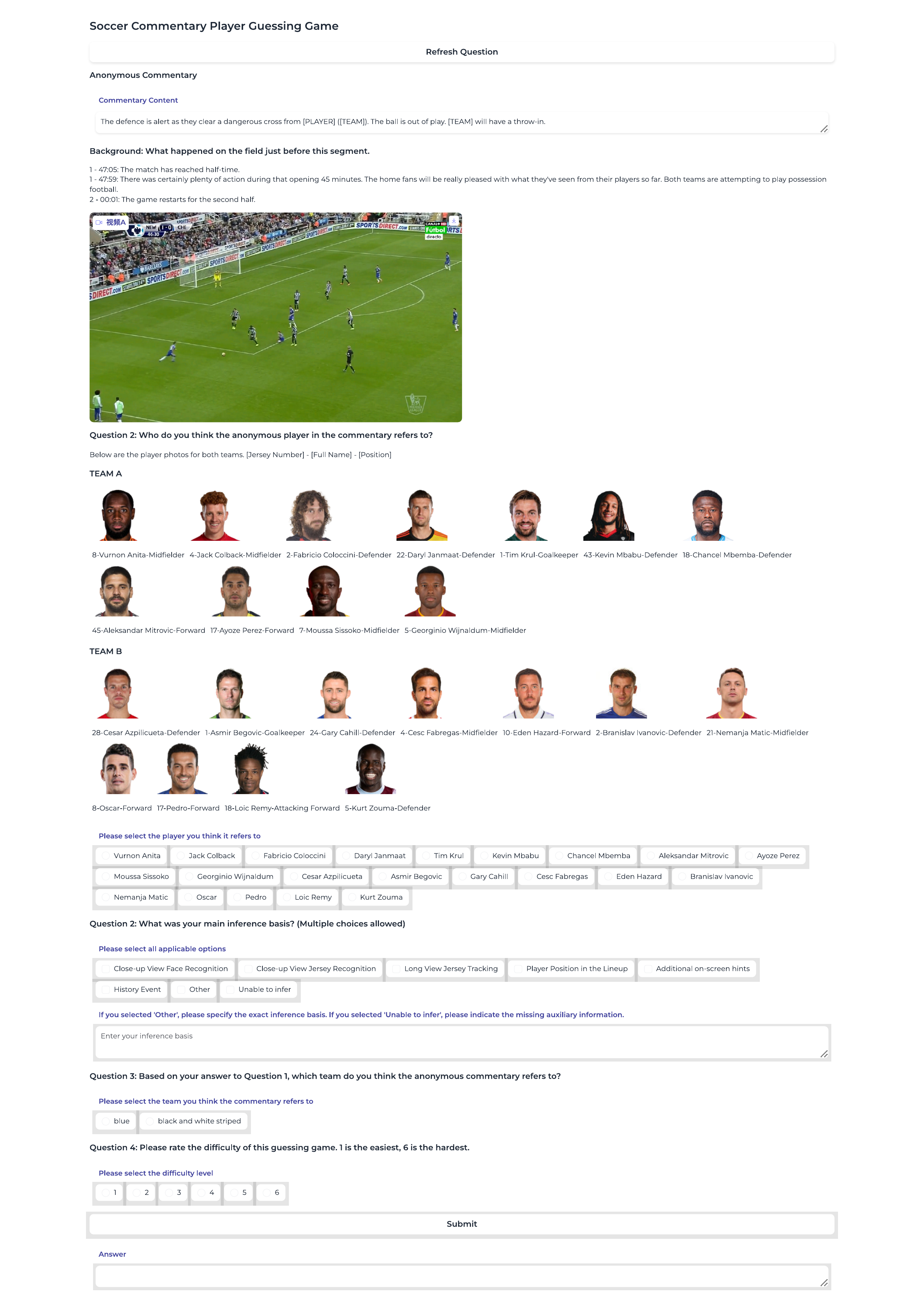}
  \captionsetup{justification=centering}
  \caption{Visual Interface of the `Player Guessing' Game.}
  \label{fig:Interface}
\end{figure*}

The visual interface of the ``Player Guessing'' game in the empirical study is shown in Fig.~\ref{fig:Interface}.
We exclude the visual clue $Scene$ and video grounding guidance $\mathcal{W}$ but directly provide the player face images and the identity information, since human masters the video grounding and fine-grained jersey tracking tasks.

\subsection{View Classification Prompt~(Sec.~IV-C)}
\begin{promptbox}
You are a professional soccer game commentator.
Shot views in a soccer video are generally categorized into four classes: long view, medium view, close-up view, and out-of-field view.
Please classify the input image and directly output the class label. DO NOT add any explanation.
\end{promptbox}

\subsection{Player Jersey Recognition Prompt~(Sec.~IV-C)}
\begin{promptbox}
Please describe the details in the video briefly. Put quality before quantity.
Details includes: 1. the time and score shows at the scoreboard.
2. the name, jersey color, jersey number and the actions of each player ONLY IF it is clear without blur.
\end{promptbox}
                 
\subsection{Answer-Guided Reasoning Prompt~(Sec.~IV-E)}

Query about [Team]:

\begin{promptbox}
You are a professional soccer game commentator. Here is the anonymized soccer commentary for this 30 seconds broadcast soccer game video: \{$Commentary$\} It describes a ``\{$Label$\}'' type of event during the soccer game between the home team: \{$Team_h$\}~(\{$Color_h$\} jersey) and the away team: \{$Team_a$\}~(\{$Color_a$\} jersey).\\
The question is: player from WHICH team conduct the event described in the anonymized commentary?\\
The commentary event is coarsely grounded at around \{$\mathcal{W}_{top5}$\} seconds in the video.\\
The correct answer should be: $\{Team_g\}$.\\
Now, craft a brief and cohesive reasoning path that deduces this team based on the visual and context clues provided. Begin your reasoning without revealing that you know the answer to this team, using a tone of exploration and inference.
Please break down the task by two steps to find out all the correct the clues, allow the observations to naturally lead you to the correct team, enhancing the accuracy of your deductions. Make sure you deduce to the correct answer: $\{Team_g\}$.\\
Step 1: Video Grounding. Carefully analyze which scene is the most related to the commentary event, where player from $\{Team_g\}$ is experiencing such event.\\
Step 2: Analyze the Attack and Defense Sides. Observe the attack and defense sides around the grounded view. Given the grounded view, SECRETLY discuss why player from $\{Team_g\}$ is more likely to conduct the mentioned type of event.\\
Please only generate the valid clues. DO NOT directly mention $\{Team_g\}$ as your reasoning intention. First generate your procedure of reasoning, and finally output the grounding segment and the correct answer in the JSON format of $\{Example\}$.
\end{promptbox}

Query about [Player]:
\begin{promptbox}
You are a professional soccer game commentator. Here is the anonymized soccer commentary for this 30 seconds broadcast soccer game video: \{$Commentary$\} It describes a ``\{$Label$\}'' type of event during the soccer game between the home team: \{$Team_h$\}~(\{$Color_h$\} jersey) and the away team: \{$Team_a$\}~(\{$Color_a$\} jersey).\\
The question is: Who is the anonymized [PLAYER] mentioned in the commentary?\\ 
The commentary event is coarsely grounded at around \{$\mathcal{W}_{top5}$\} seconds in the video.
The lineup of the home team is: \{$\mathcal{T}_h(t)$\} and the lineup of the away team is \{$\mathcal{T}_a(t)$\}.
The history timeline (recent happenings on the court) is: {history timeline}
The coarse scene captioning for the views in this 30 seconds video are:
\{$Scene(t)$\}\\
The correct answer should be player $\{Player_g\}$ from $\{Team_g\}$. Bear in mind his hash: $\{Player_g\}$, his name: $\{Team_g\}$, his jersey number: $\{Number_g\}$ and his role: $\{Role_g\}$.\\
Now, craft a brief and cohesive reasoning path that deduces this player based on the visual and context clues provided. Begin your reasoning without revealing that you know the answer to this player or his team, using a tone of exploration and inference.
Please firstly summarize the anonymized player you are looking for from the commentary, then break down the task step by step to find out all the correct the clues, allow the observations to naturally lead you to the correct player, enhancing the accuracy of your deductions. Make sure you deduce to the correct answer: $\{Player_g\}$.\\
Step 1: Video Grounding. Analyze which scene is the most related to the commentary event.\\
Step 2: Analyze the Attack and Defense Sides. Observe the attack and defense sides around the grounded view. Given the grounded view, SECRETLY discuss why player from $\{Team_g\}$ is more likely to conduct the mentioned type of event.\\
Step 3: Identify the Player Involved. Observe the appeared players in \{$\mathcal{W}_{top5}$\} seconds of the video. SECRETLY detect whether $\{Player_g\}$ is near the soccer ball in the grounded view, whether his visual identity clue including the jersey number and name could be observed from a clear view. Also, SECRETLY observe whether the jersey number, name or face of $\{Player_g\}$ is caught in any of the close-up view. Because in the broadcast footage of football matches, player $\{Player_g\}$ is likely to appear in the close-up views to show his relationship with the event, but the close-up view may not be grounded to the event. Ignore the irrelevant close-up view for other players.\\
Step 4: Context knowledge Analysis. SECRETLY search the history timeline to find any potential evidence to support the reasoning. The role of player $\{Player_g\}$ may also have connection with the mentioned type of event.\\
Please only generate the valid clues. Do not directly mention the correct answer in your reasoning intention. First generate your procedure of reasoning, and finally output the grounding segment and the correct answer in the JSON format of $\{Example\}$.
\end{promptbox}

\subsection{SFT Reasoning Prompt~(Exp.~3)}

Query about $Player$:

\begin{promptbox}
...(Similar to the above one)...\\
\strike{Now, craft a brief and cohesive reasoning path that }\\
\strike{deduces this player ...}\\
Please firstly summarize the anonymized player you are looking for from the commentary, then break down the task step by step to find out the answer.\\
Step 1: Video Grounding. Analyze which scene is the most related to the commentary event.\\
Step 2: Analyze the Attack and Defense Sides. Observe the attack and defense sides around the grounded view. Given the grounded view, discuss which side is more likely to conduct mentioned type of event.\\
Step 3: Identify the Player Involved. Observe the appeared players in \{$\mathcal{W}_{top5}$\} seconds of the video. Detect the player near the soccer ball in the grounded video. Observe his visual identity clue including the jersey number and name from a clear view. Also, pay attention to the jersey number, name or face appeared in the close-up view. Because in the broadcast footage of football matches, the [PLAYER] might appears in the close-on views to show his relationship to the event, although the close-on view may not be directly grounded as the event.\\
Step 4: Context knowledge Analysis. Combine the history timeline to find any potential evidence to support the reasoning. Also, take into consideration which roles may have connection with the mentioned type of event.\\
Please first generate your procedure of reasoning, and finally output the grounding segment and the answer in the JSON format of $\{Example\}$.
\end{promptbox}

Query about $Player_c$:

\begin{promptbox}
...(Similar to the above one)...\\
\strike{Who is the anonymized [PLAYER] mentioned in the}\\
\strike{commentary?}\\
Who is the anonymized [PLAYER] from $\{Team_g\}$ mentioned in the commentary?\\
...(Similar to the above one)...
\end{promptbox}

\subsection{GRPO Reasoning Prompt~(Exp.~3)}

\begin{promptbox}
...(Similar to the above one)...\\
\strike{Please first generate your procedure of reasoning,}\\
\strike{and finally output the grounding segment and the}\\
\strike{answer in the JSON format of $\{Example\}$.}\\
Output the thinking process in \textless{}think\textgreater{}
 \textless{}/think\textgreater{}
 and final answer in \textless{}answer\textgreater{}
 \textless{}/answer\textgreater{} tags, i.e., \textless{}think\textgreater{} reasoning process here \textless{}/think\textgreater{} \textless{}answer\textgreater{} answer here \textless{}/answer\textgreater{}. \\
(If Query about $Team$:)\\
A. hometeam\quad B. awayteam \\
(If Query about $Player$:)\\
A.~\{Hash~A\}\quad B.~\{Hash~B\} \quad ...\quad U.~\{Hash~U\} \quad V.~\{Hash~V\}
\end{promptbox}

\section{Implementation Details in Stage II}

\subsection{Ablation study on $k$ in history context length}

We currently adopt $k=1$ in $HistoryEvent$ timeline based on the Stage I ablation (Appendix~\ref{Details of the Fine-grained Shot Analysis}.5), where redundant history events ($k>1$) were found to hurt entity alignment as they bring distractions for reasoning.

As to the internal context of Stage II, besides the full $KeyEvent$ timeline of the current type of event, we also evaluate \textsc{GameSight} with varying $k$ lengths in terms of discourse cohesion and text similarity.

\begin{table}[ht]
\centering
\caption{Evaluation results across different $k$ values.}
\label{table_k_values}
\resizebox{\columnwidth}{!}{
\begin{tabular}{lcccccc}
\toprule
 & \textbf{$k$=0} & \textbf{$k$=1} & \textbf{$k$=3} & \textbf{$k$=5} & \textbf{$k$=10} & \textbf{All} \\
\midrule
\textbf{Deep Cohesion} & 0.804 & \underline{0.956} & \textbf{0.960} & 0.540 & 0.620 & 0.630 \\
\textbf{BLEU} & 17.020 & \textbf{17.409} & 16.859 & \underline{17.060} & 16.812 & 16.892 \\
\textbf{METEOR} & 8.644 & \textbf{9.051} & 8.815 & \underline{8.928} & 8.644 & 8.827 \\
\textbf{ROUGE-L} & \textbf{10.556} & \underline{10.497} & 10.181 & 10.356 & 10.323 & 10.346 \\
\textbf{CIDEr} & 2.780 & \underline{3.442} & 2.326 & 3.025 & 3.306 & \textbf{3.530} \\
\bottomrule
\end{tabular}
}
\end{table}

The results show that k=1 has the best overall performance, indicating that immediate context is most helpful for coherent commentary, which is similar to the conclusion in Stage I. Larger k tends to introduce noise, reflecting current limitations in handling long-term context effectively.

\subsection{Knowledge Extraction Prompt~(Sec.~V-A)}
\begin{promptbox}
You are a professional soccer game commentator. Here is a transcription of soccer commentary: \{$Transcription$\}
It comes from the soccer game between team: \{$Teams$\}, league: \{$League$\}, season: \{$Season$\}, date and time: \{$Date$\}.
Now answer the question step by step: \\
Step 1. Analyze if any statistics is used in the commentary. If no statistics, return "None". Else, describe the statistic facts in a sentence.\\
Step 2. Generate the related question specifically to which the answer should be the statistics used in the commentary. If no statistics, return "None". Else, generate the related question.\\
Please provide the answer in the following JSON format: \{$Example$\}.
\end{promptbox}

\subsection{Commentary Refinement Prompt~(Sec.~V-D)}

Generate questions related to external statistics:
\begin{promptbox}
You are a professional soccer game commentator.
Here is a soccer commentary with raw description: \{$Commentary$\}\
It comes from the soccer game between home team: \{$Team_h$\} and away team: \{$Team_a$\} at league: \{$League$\}, season: \{$Season$\}, date and time: \{$Date$\}.
Since the raw description only describes the current happening without any knowledge of history statistics (knowledge beyond this current game), please generate some query that you want to know before you refine the commentary. I will show you the answer from the external database. Use accurate reference of time, date, player full name and team name in the query. Do not ask abstract question.\\
For example, instead of asking "How has <player> performed" or "What impact did <player> have", ask "How many goals has <player> scored in season \{$Season$\} before \{$Date$\}?" Instead of asking "What are the current statistics for <team>", ask "List all games that played against\{$Teams$\} in season \{$Season$\} and league \{$League$\}".\\
Please generate four questions in the following JSON format: \{$Example$\}.
\end{promptbox}

Refine commentary with external and internal knowledge:
\begin{promptbox}
You are a professional soccer game commentator.
Here is a soccer commentary with raw description: \{$Commentary$\}\
It comes from the soccer game between home team: \{$Team_h$\} and away team: \{$Team_a$\} at league: \{$League$\}, season: \{$Season$\}, date and time: \{$Date$\}.
It describes a "\{$Label$\}" type of event at the game time of "\{$GameTime$\}".\\
Since the raw description only describes the current happening without any knowledge of history game context and season statistics, you are provided with the following knowledge that you may need to further refine the commentary.\\
External Knowledge: \{$External~Knowledge$\} \\
Internal Knowledge: \{$Internal~Knowledge$\} \\
Now, refine the raw commentary in a professional style. You are contributing to only a segment for the commentary of the whole game, so do not be verbose, and decrease the reference to the static information throughout the whole game, because it could have been mentioned previously. Only pick up the unique, precise and necessary knowledge to fulfill your professional commentary. Please directly output your refinement.
\end{promptbox}

\section{Implementation Details in Experiments}
\label{Implementation Details in Experiments}
In both stages, we used the original anonymous commentary from the MatchTime~(SN-caption-test) dataset for \textsc{GameSight} inference. It is because its ground-truth entities are available in the corresponding original non-anonymous commentary, which naturally form up the proposed data pairs. Meanwhile, it represents the upper bound of the framework's prior knowledge as input.

In Sec~6.2, commentary are evaluated from two aspects: 

\noindent(1) the accuracy of knowledge reference, where accurate entity matters (Exp. 4); 

\noindent(2) the general quality in how `human centric' it is, where the concrete reference of any exact player/team is not the primary focus (Exp. 3,5,6). 

\subsection{Test Sets}
We employed two test sets to conduct the evaluation.

\noindent\textbf{1.~Live televised commentary test set}~(Exp. 3,5,6)
We curated a new Live-televised commentary test set from Goal~(CITK 2023), which originally contains human proofread ASR transcriptions for 20 matches from the SoccerNet-v2 dataset. We take the three overlapped matches between Goal and SN-caption-test-align. The other 18 matches are excluded because they are part of the training set of the conventional models. To obtain the ground truth live-televised commentary of the 30 seconds video segment that conventional models take as input, for each event timestamps, we used GPT-4o-mini to extract the related commentary from the ASR transcription within a 60 seconds context window.
To evaluate the overall quality of different commentary, all commentary are anonymized in this experiment.

\noindent\textbf{2.~External knowledge accuracy test set}~(Exp. 4)

Catering to the evaluation for external knowledge accuracy and robustness, we manually tailor a question set with 100 external statistical questions (50 for players, 50 for teams). 
While involving complex scenarios, the statistical knowledge relevant to commentary primarily revolves around a finite set of structured relationships among teams, players, events, and scores, as demonstrated below. Given this underlying regularity, the key challenge for the SoccerKAG system is the accurate translation of diverse natural language queries into correct schema and SQL representations (NL $\to$ SQL). 
Thereby, we generated five alternate phrasings for each question using GPT-4o, forming up 500 questions covering human commentators' reference habits. 
We elaborate 30 question samples from the external statistical question set.

\begin{promptbox}
Question 1: How many goals has Alexis Sanchez scored in season 2016-2017 before 2016-11-22?

Question 2: How many goals has Cristiano Ronaldo scored in season 2014-2015 before 2015-04-17?

Question 3: How many goals has Sergio Aguero scored for Manchester City in the 2015-2016 season before 2015-08-15?

Question 4: How many goals has Cesc Fabregas scored for Chelsea in the 2015-2016 season before 2015-10-02?

Question 5: How many goals has Lionel Messi scored in season 2016-2017 before 2017-04-25?

Question 6: How many yellow cards has Marco Verratti received in the 2016-2017 UEFA Champions League season before 2016-11-22?

Question 7: How many yellows cards has Cesc Fabregas received in the 2015-2016 season before 2015-10-23?

Question 8: How many yellow cards has Leon Goretzka received in the 2015-2016 Bundesliga season before 2015-11-07?

Question 9: How many yellow cards has Jordan Henderson received in the 2016-2017 season before the match on 2016-08-26?

Question 10: How many assists did Toni Kroos provide for Real Madrid in the UEFA Champions League during the 2016-2017 season before 2017-05-01?

Question 11: How many assists did Jesus Navas record for Manchester City in the Premier League during the 2015-2016 season before 2015-08-15?

Question 12: How many fouls has Dejan Lovren committed in the 2016-2017 season before 2016-08-26?

Question 13: How many fouls has Daniel Carvajal committed in the 2015-2016 UEFA Champions League season before 2015-11-24?

Question 14: What has been Kurt Zouma's disciplinary record in the 2015-2016 season before 2016-02-02?

Question 15: How many cards did Gary Cahill receive in season 2015-2016 before the match on 2015-10-02?

Question 16: How many penalties did Bayern Munich receive in the season 2015-2016 before the game on 2015-10-23?

Question 17: How many penalties has West Brom been awarded in the season 2015-2016 before 2015-08-22?

Question 18: How many corners have Atl. Madrid taken in the UEFA Champions League during the 2016-2017 season before 2017-05-01?

Question 19: What was Paris SG's record in the UEFA Champions League up to the date of 2016-11-22 during the 2016-2017 season?

Question 20: How many games has Arsenal played in the UEFA Champions League during the 2016-2017 season before 2016-11-22?

Question 21: What were Bayern Munich's overall statistics in the 2015-2016 Bundesliga season before the match against FC Koln on 2015-10-23?

Question 22: How many wins did Paris SG have in the 2016-2017 UEFA Champions League before their match against Arsenal on 2016-11-22?

Question 23: What were the results of the last three matches played by Paris SG in the 2016-2017 UEFA Champions League before March 7, 2017?

Question 24: What were the results of Barcelona's previous five matches in the UEFA Champions League during the 2016-2017 season before the game against Paris SG on March 7, 2017?

Question 25: How many goals has Barcelona scored in the 2014-2015 La Liga season before the match on 2015-04-24?

Question 26: How many goals has Napoli scored in the UEFA Champions League during the 2016-2017 season before 2016-11-01?

Question 27: List all yellow cards received by Real Madrid players in the UEFA Champions League season 2014-2015 up to 2015-03-09.

Question 28: List all free kicks awarded to West Ham in their matches within the 2015-2016 season before 2015-10-23.

Question 29: What has been Chelsea's away performance in the Premier League during the season 2015-2016 prior to the match on 2016-02-02?

Question 30: List all matches FC Porto has played in the UEFA Champions League season 2014-2015 before 2015-04-14.

\end{promptbox}

\subsection{User Study~(Exp.~5)}

We invited 12 participants with basic soccer knowledge (aged 19 to 26) to participate in the user study. Participants are asked to watch ten silent soccer clips sampled from the overlapped matches between Goal and SN-Caption-test-align, then score 1 to 5 for the quality of five pieces of corresponding commentary. 

\subsection{Commentary Classification Prompt~(Exp.~6)}

\begin{promptbox}
There are \{$k$\} sentences in this segment of soccer commentary. Classify each sentence into one of three category: "Description", "Explanation", and "Comment", under the followed instruction: \\
Description: Real-time narration of on-screen events, focusing on immediate actions, and technical movements. \\
Explanation: Contextual analysis or background information that adds depth to the game, including details like player formations, weather conditions, or game rules for referee decisions.\\
Commentary: Expressive and evaluative remarks that summarize or highlight critical moments, often adding emotional weight or perspective. \\
While categorizing, focus on the dominant intent of the statement but flexible combining categories in a single sentence is allowed. 
Directly output the list of decisions on each sentence without repeating the sentence, for instance: 1.Description 2.Commentary 3.Description and Explanation.
Here is the text to be classified: \{$Commentary$\}.
\end{promptbox}

\section{Case Study}

\subsection{Case Study in Visual Reasoning}

Please refer to the good cases and bad cases shown in Fig.~11-Fig.~14, or the video versions on our demo page.
\begin{figure*}
  \includegraphics[width=\linewidth]{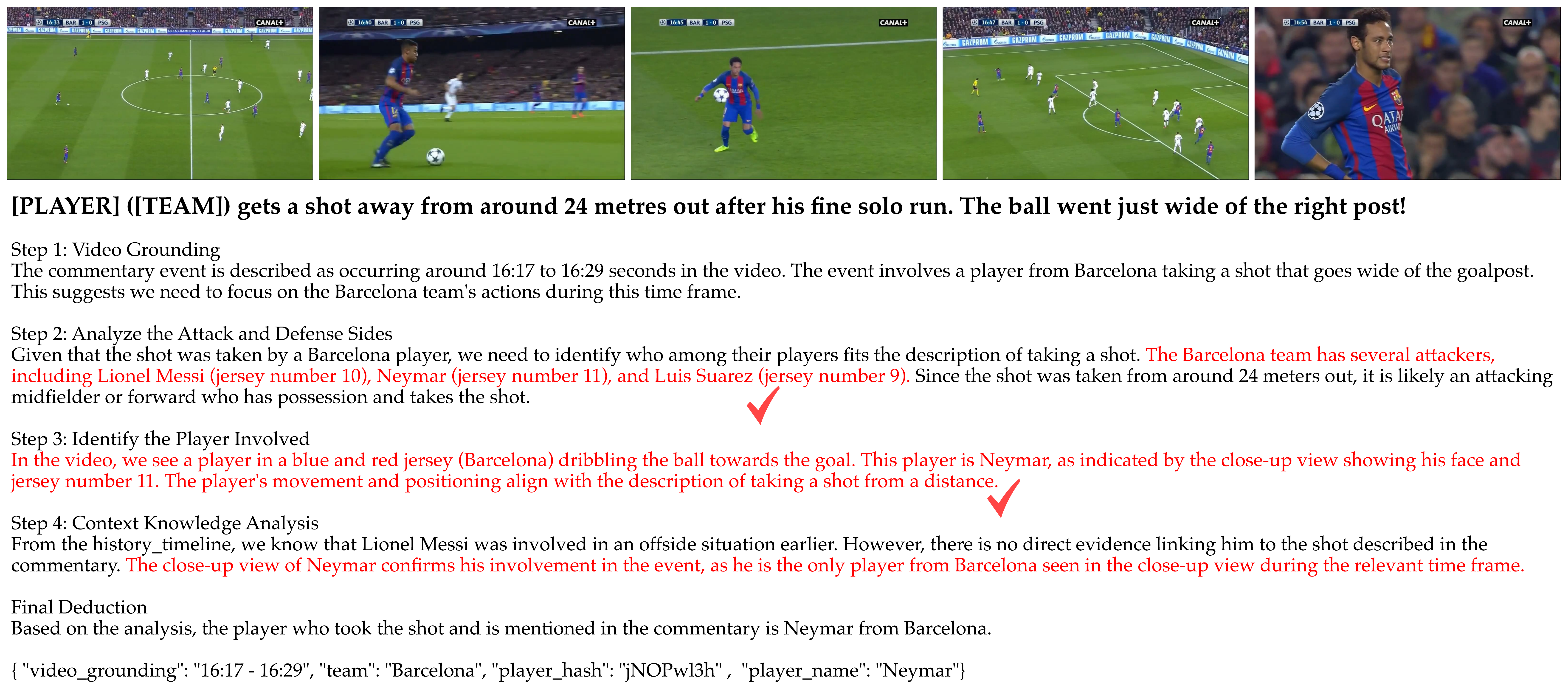}
  \caption{Good case by analyzing close view recognized face with player role.}
\end{figure*}
\begin{figure*}
  \includegraphics[width=\linewidth]{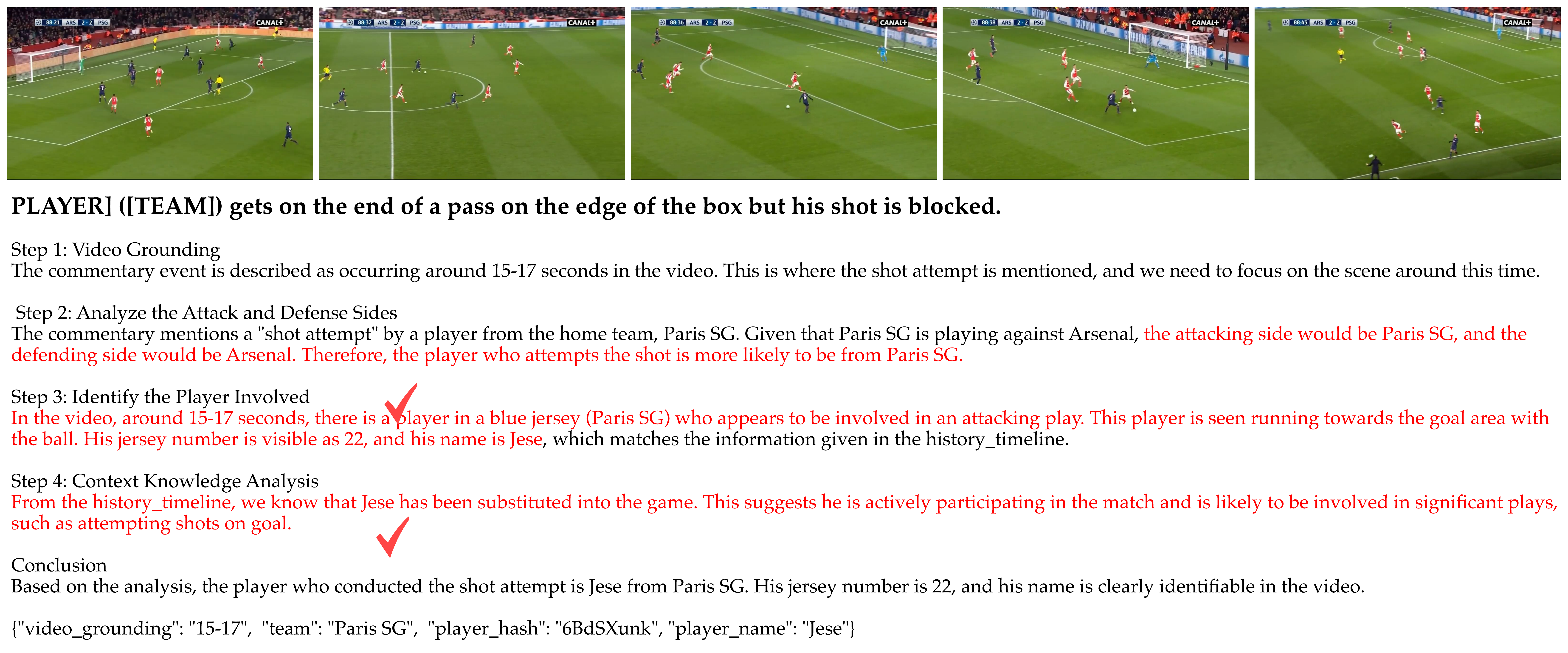}
  \caption{Good case by observation in the long view and analyzing history timeline.}
\end{figure*}
\begin{figure*}
  \includegraphics[width=\linewidth]{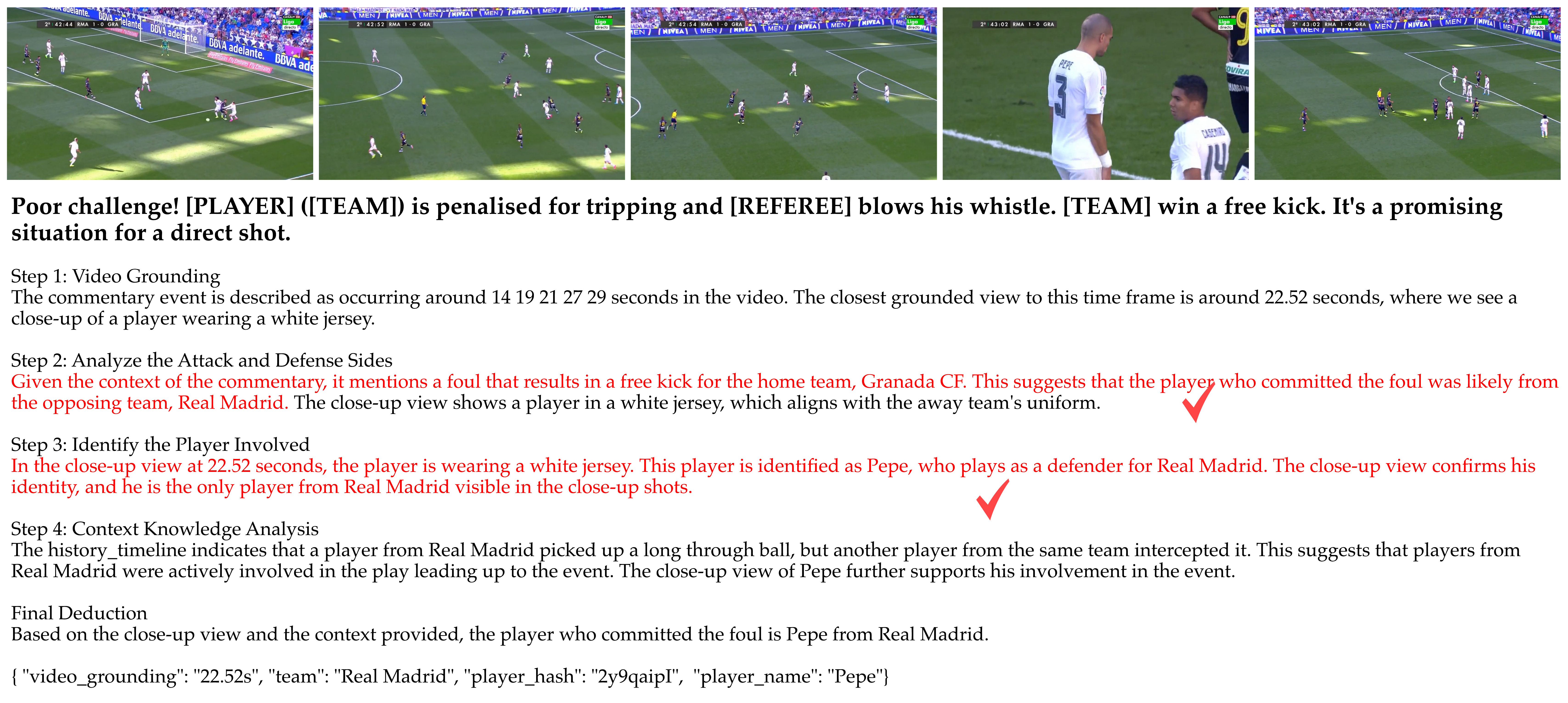}
  \caption{Good case by analyzing close view recognized face with player role.}
\end{figure*}
\begin{figure*}
  \includegraphics[width=\linewidth]{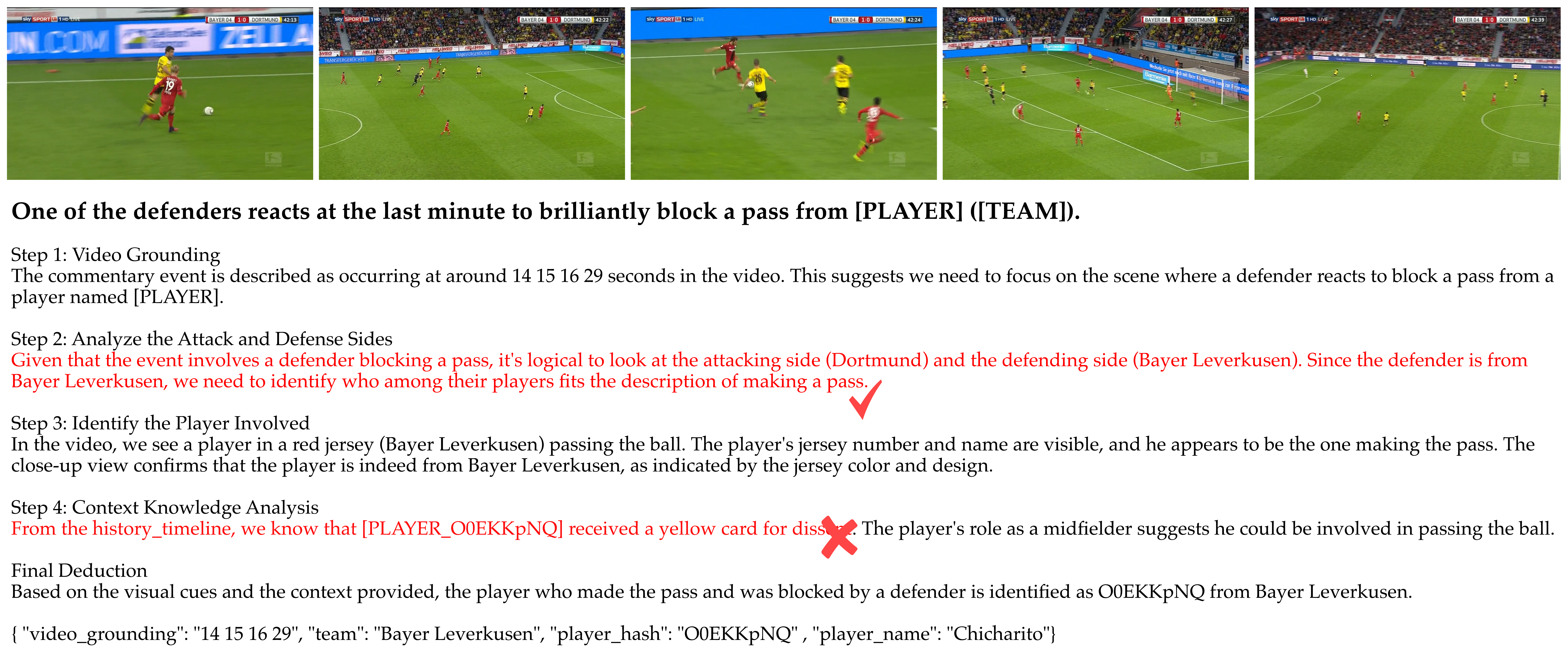}
  \caption{Bad case with wrong deduction.}
\end{figure*}

\subsection{Case Study in Commentary Refinement}

Good Cases:

\begin{promptbox}
\textbf{MatchTime}: Edinson Cavani (Paris SG) times his run too early and the linesman raises his flag for offside.

\textbf{\textsc{GameSight}}: In the 52nd minute of the match, Edinson Cavani times his run too early, and the linesman raises his flag for offside. It's a frustrating moment for the Paris SG forward, who has already found the net twice in this UEFA Champions League season. The away team will be hoping to capitalize on their attacking efforts as they trail 2-1 against Arsenal here at the Emirates Stadium.
\end{promptbox}

\begin{promptbox}
\textbf{MatchTime}: Lucas (Paris SG) takes a mid-range free kick and his stunning strike crashes against the crossbar.

\textbf{\textsc{GameSight}}: Lucas Moura steps up for a mid-range free kick, showcasing his trademark technique! He strikes it brilliantly, but it crashes against the crossbar, sending shockwaves through the Emirates Stadium. Paris SG is putting the pressure on Arsenal, as Lucas looks to add to his three goals this season. The score remains 1-1 here in the second half!
\end{promptbox}

\begin{promptbox}
\textbf{MatchTime}: Willian (Chelsea) takes the resulting corner which is well defended. Chelsea will take a corner kick.

\textbf{\textsc{GameSight}}: Willian steps up to take the corner for Chelsea after a well-defended sequence. They've been pressing hard, looking to extend their lead. The current score stands at 2-1 in favor of the visitors. The last corner was also for Chelsea, but they’ll look to make this one count at Carrow Road. Let’s see how Norwich's defense responds!
\end{promptbox}

\begin{promptbox}
\textbf{MatchTime}: Grzegorz Krychowiak (Sevilla) brings an opponent down and Svein Moen interrupts the game.

\textbf{\textsc{GameSight}}: Grzegorz Krychowiak brings down an opponent, and referee Svein Moen swiftly halts play. It's a strategic foul from the Sevilla midfielder as they look to regain control of the game. We're now at 33 minutes and 38 seconds into the second half, with Sevilla trailing 1-3 against Manchester City. Krychowiak, who has already received a yellow card earlier, will need to be cautious for the remainder of the match.
\end{promptbox}



\begin{promptbox}
\textbf{MatchTime}: Ramires (Chelsea) fouls his opponent and receives a yellow card from the referee, but he is clearly not happy with the decision.

\textbf{\textsc{GameSight}}: As we approach the closing stages of the first half, Ramires from Chelsea commits a foul, much to his dismay, and is issued a yellow card by the referee. The Brazilian midfielder clearly disagrees with the decision as he gestures his frustration. Meanwhile, Southampton has already received two yellow cards in this match, with Ryan Bertrand and Sadio Mane already cautioned. The tension is palpable as Chelsea leads 1-0 following Willian's early free-kick goal.
\end{promptbox}

\begin{promptbox}
\textbf{MatchTime}: Franck Ribery (Bayern Munich) is shown a yellow card by the referee.

\textbf{\textsc{GameSight}}: As we approach the midway point of the second half, Franck Ribery has just received a yellow card for Bayern Munich after a late challenge. This adds to the growing tension in this highly contested match at Signal Iduna Park, with Dortmund leading 1-0 following Pierre-Emerick Aubameyang's opening goal in the first half. Ribery's booking could have significant implications as both teams battle for control. This marks the third yellow card of the match, with Adrian Ramos and Marc Bartra already cautioned for Dortmund. The stakes are high as we inch closer to the final whistle.
\end{promptbox}

\begin{promptbox}
\textbf{MatchTime}: Juanmi (Malaga) scores a goal! A perfectly produced cross puts him one-on-one with the goalkeeper and he plants a brilliant header down in the middle of the net, leaving the goalkeeper beaten!

\textbf{\textsc{GameSight}}: Juanmi has found the back of the net for Malaga! At the 70th minute, a perfectly timed cross allows him to break free one-on-one with Iker Casillas. The young forward rises above the defenders to deliver a brilliant header down into the center of the goal, leaving the Real Madrid keeper with no chance. This goal comes as Malaga seeks to fight back after being two goals down. The score now stands at 2-1, and the home crowd at the Santiago Bernabéu is stunned!
\end{promptbox}

Bad Cases:

\begin{promptbox}
\textbf{MatchTime}: Matthias Ginter (Dortmund) can't expect anything else than a yellow card for his hard tackle. And a yellow it is. Manuel Grafe didn't think twice about pulling it out of his pocket.

\textbf{\textsc{GameSight}}: Matthias Ginter of Dortmund has been booked, receiving a yellow card for a robust challenge. Referee Manuel Grafe didn't hesitate to brandish the card, a clear indication of the game's growing physicality. This marks another disciplinary action in what has been a fiercely contested encounter, \textbf{with Leverkusen already having three yellow cards to their name}. The score remains 1-0 to Bayer Leverkusen as we approach the half-time mark. \textbf{(Forth yellow cards)}
\end{promptbox}

\begin{promptbox}
\textbf{MatchTime}: The game is interrupted as Cristiano Ronaldo (Real Madrid) is flagged offside.

\textbf{\textsc{GameSight}}: The game takes a brief pause as Cristiano Ronaldo is caught offside, just as Real Madrid was looking to capitalize on their significant lead here at the Santiago Bernabéu. With the score at 4-1 and about a minute to go until halftime, the home side has controlled the play and \textbf{maintains 62\% possession}. This offside decision serves as a reminder for Madrid to be mindful of their positioning as they continue to push for more goals. \textbf{(Hallucination, we don't have ball possession information.)}
\end{promptbox}

\begin{promptbox}
\textbf{MatchTime}: Arjen Robben (Bayern Munich) delivers a low pass into the box which is intercepted. The referee and his assistant both point at the corner flag. Bayern Munich will have an opportunity to threaten the opposition's goal.

\textbf{\textsc{GameSight}}: Arjen Robben weaves into the box and delivers a low pass towards the goal, but it’s intercepted by the FC Köln defense. However, the referee swiftly points to the corner flag—Bayern Munich will have a corner kick, \textbf{their fifth of the match so far}. With the score at 1-0, Bayern is looking to further capitalize on their attacking momentum here at the Allianz Arena. \textbf{(Sixth corner kicks.)}
\end{promptbox}

\begin{promptbox}
\textbf{MatchTime}: Goal! Neymar (Barcelona) slams home the penalty with a precise shot which goes inside the right post.

\textbf{\textsc{GameSight}}: And there it is! Goal! Neymar steps up to take the penalty and finishes with tremendous precision, slamming the ball home just inside the right post. That's a crucial strike for Barcelona as they extend their lead to 4-1 against Paris SG. The home crowd at Camp Nou erupts! This is Neymar's second goal of the match, following his earlier stunning free kick. Just what Barcelona needed as they push for a commanding victory in this Champions League clash! \textbf{(Score should be updated to 5-1.)}
\end{promptbox}

%% file: icme2026references.bib
@report{fifa_football_landscape_2021,
  title       = {The Football Landscape},
  author      = {{FIFA}},
  year        = {2021},
  institution = {FIFA},
  url         = {https://publications.fifa.com/en/vision-report-2021/the-football-landscape/},
}

@inproceedings{Somers2024SoccerNetGameState,
    title = {{SoccerNet} Game State Reconstruction: End-to-End Athlete Tracking and Identification on a Minimap},
    author = {Somers, Vladimir and Joos, Victor and Giancola, Silvio and Cioppa, Anthony and Ghasemzadeh, Seyed Abolfazl},
    booktitle = cvsports,
    month = Jun,
    year = {2024},
    address = city-seattle
}

@inproceedings{banerjee2005meteor,
  title={METEOR: An automatic metric for MT evaluation with improved correlation with human judgments},
  author={Banerjee, Satanjeev},
  booktitle={Proceedings of the acl workshop on intrinsic and extrinsic evaluation measures for machine translation and/or summarization},
  pages={65--72},
  year={2005}
}

@inproceedings{vedantam2015cider,
  title={Cider: Consensus-based image description evaluation},
  author={Vedantam, Ramakrishna},
  booktitle={Proceedings of the IEEE conference on computer vision and pattern recognition},
  pages={4566--4575},
  year={2015}
}

@inproceedings{papineni2002bleu,
  title={Bleu: a method for automatic evaluation of machine translation},
  author={Papineni, Kishore and Roukos, Salim and Ward, Todd and Zhu, Wei-Jing},
  booktitle={Proceedings of the 40th annual meeting of the Association for Computational Linguistics},
  pages={311--318},
  year={2002}
}

@book{hedrick2000art,
  title={The art of sportscasting: How to build a successful career},
  author={Hedrick, Tom},
  year={2000},
  publisher={Taylor Trade Publications}
}

@article{zhang2017shanghai,
  author  = {Zhang, Desheng and Li, Feng and Wang, Ziye},
  title   = {Three Kinds of Logic on Sport Interpretation \& Commentary and Their Scientific Application},
  journal = {Journal of Shanghai University of Sport},
  year    = {2017},
  volume  = {41},
  number  = {2},
  pages   = {15--20},
  doi     = {10.16099/j.sus.2017.02.003},
  note    = {[J]}
}

@misc{strand_soccerrag_2024,
	title = {{SoccerRAG}: Multimodal Soccer Information Retrieval via Natural Queries},
	url = {http://arxiv.org/abs/2406.01273},
	shorttitle = {{SoccerRAG}},
	number = {{arXiv}:2406.01273},
	publisher = {{arXiv}},
	author = {Strand, Aleksander Theo and Gautam, Sushant and Midoglu, Cise and Halvorsen, Pål},
	urldate = {2024-08-27},
	year = {2024},
	archivePrefix={arXiv},
	eprint = {2406.01273 [cs]},
}

@misc{deliege_soccernet-v2_2021,
	title = {{SoccerNet}-v2: A Dataset and Benchmarks for Holistic Understanding of Broadcast Soccer Videos},
	url = {http://arxiv.org/abs/2011.13367},
	shorttitle = {{SoccerNet}-v2},
	number = {{arXiv}:2011.13367},
	publisher = {{arXiv}},
    archivePrefix={arXiv},
	author = {Deliège, Adrien and Cioppa, Anthony and Giancola, Silvio},
	urldate = {2024-08-28},
	date = {2021-04-19},
    year = {2021},
	archivePrefix={arXiv},
	eprint = {2011.13367 [cs]},
}

@article{li2023videochat,
  title={Videochat: Chat-centric video understanding},
  author={Li, KunChang and He, Yinan and Wang, Yi and Li, Yizhuo and Wang, Wenhai and Luo, Ping and Wang, Yali and Wang, Limin and Qiao, Yu},
  journal={arXiv preprint arXiv:2305.06355},
  year={2023}
}

@misc{qi_goal_2023,
	title = {{GOAL}: A Challenging Knowledge-grounded Video Captioning Benchmark for Real-time Soccer Commentary Generation},
	url = {http://arxiv.org/abs/2303.14655},
	shorttitle = {{GOAL}},
	number = {{arXiv}:2303.14655},
	publisher = {{arXiv}},
    archivePrefix={arXiv},
	author = {Qi, Ji and Yu, Jifan and Tu, Teng and Gao, Kunyu and Xu, Yifan and Guan, Xinyu},
	urldate = {2024-07-19},
	year = {2023},
	archivePrefix={arXiv},
	eprint = {2303.14655 [cs]},
}

@misc{qwen2.5-VL,
    title = {Qwen2.5-VL},
    url = {https://qwenlm.github.io/blog/qwen2.5-vl/},
    author = {{Qwen Team}},
    month = {January},
    year = {2025}
}

@misc{li2024llavaonevisioneasyvisualtask,
      title={LLaVA-OneVision: Easy Visual Task Transfer}, 
      author={Bo Li and Yuanhan Zhang and Dong Guo and Renrui Zhang and Feng Li and Hao Zhang and Kaichen Zhang and Peiyuan Zhang and Yanwei Li and Ziwei Liu and Chunyuan Li},
      year={2024},
      eprint={2408.03326},
      archivePrefix={arXiv},
      primaryClass={cs.CV},
      url={https://arxiv.org/abs/2408.03326}, 
}

@misc{hu2021loralowrankadaptationlarge,
      title={LoRA: Low-Rank Adaptation of Large Language Models}, 
      author={Edward J. Hu and Yelong Shen and Phillip Wallis and Zeyuan Allen-Zhu and Yuanzhi Li and Shean Wang and Lu Wang and Weizhu Chen},
      year={2021},
      eprint={2106.09685},
      archivePrefix={arXiv},
      primaryClass={cs.CL},
      url={https://arxiv.org/abs/2106.09685}, 
}

@techreport{lin2004rouge,
  title={{ROUGE}: A Package for Automatic Evaluation of Summaries},
  author={Lin, Chin-Yew},
  year={2004},
  institution={ACL Workshop on Text Summarization Branches Out},
  number={W04-1013},
  note={https://aclanthology.org/W04-1013}
}

@article{rao2025socceragent,
        title={Multi-Agent System for Comprehensive Soccer Understanding},
        author={Rao, Jiayuan and Li, Zifeng and Wu, Haoning and Zhang, Ya and Wang, Yanfeng and Xie, Weidi},
        journal={arXiv preprint arXiv:2505.03735},
        year={2025}
}

@misc{jiang2025domainadaptationvlmsoccer,
      title={Domain Adaptation of VLM for Soccer Video Understanding}, 
      author={Tiancheng Jiang and Henry Wang and Md Sirajus Salekin and Parmida Atighehchian and Shinan Zhang},
      year={2025},
      eprint={2505.13860},
      archivePrefix={arXiv},
      primaryClass={cs.CV},
      url={https://arxiv.org/abs/2505.13860}, 
}

@misc{you2025timesoccerendtoendmultimodallarge,
      title={TimeSoccer: An End-to-End Multimodal Large Language Model for Soccer Commentary Generation}, 
      author={Ling You and Wenxuan Huang and Xinni Xie and Xiangyi Wei and Bangyan Li and Shaohui Lin and Yang Li and Changbo Wang},
      year={2025},
      eprint={2504.17365},
      archivePrefix={arXiv},
      primaryClass={cs.CV},
      url={https://arxiv.org/abs/2504.17365}, 
}

@misc{zhu2025internvl3exploringadvancedtraining,
      title={InternVL3: Exploring Advanced Training and Test-Time Recipes for Open-Source Multimodal Models}, 
      author={Jinguo Zhu and Weiyun Wang and Zhe Chen and Zhaoyang Liu and Shenglong Ye},
      year={2025},
      eprint={2504.10479},
      archivePrefix={arXiv},
      primaryClass={cs.CV},
      url={https://arxiv.org/abs/2504.10479}, 
}

@misc{zhang2025videollama3frontiermultimodal,
      title={VideoLLaMA 3: Frontier Multimodal Foundation Models for Image and Video Understanding}, 
      author={Boqiang Zhang and Kehan Li and Zesen Cheng and Zhiqiang Hu and Yuqian Yuan},
      year={2025},
      eprint={2501.13106},
      archivePrefix={arXiv},
      primaryClass={cs.CV},
      url={https://arxiv.org/abs/2501.13106}, 
}

@inproceedings{hutto2014vader,
  title={VADER: A Parsimonious Rule-based Model for Sentiment Analysis of Social Media Text},
  author={Hutto, C.J. and Gilbert, E.},
  booktitle={Proceedings of the Eighth International Conference on Weblogs and Social Media},
  year={2014},
  url={https://www.aclweb.org/anthology/W14-3213.pdf}
}

@misc{wang-2025-open-r1-video,
  author = {Xiaodong Wang and Peixi Peng},
  title = {Open-R1-Video},
  year = {2025},
  publisher = {GitHub},
  journal = {GitHub repository},
  howpublished = {\url{https://github.com/Wang-Xiaodong1899/Open-R1-Video}}
}

@book{mcnamara2014automated,
  title={Automated evaluation of text and discourse with Coh-Metrix},
  author={McNamara, Danielle S and Graesser, Arthur C and McCarthy, Philip M and Cai, Zhiqiang},
  year={2014},
  publisher={Cambridge University Press}
}

@misc{gautam_soccernet-echoes_2024,
	title = {{SoccerNet}-Echoes: A Soccer Game Audio Commentary Dataset},
	url = {http://arxiv.org/abs/2405.07354},
	shorttitle = {{SoccerNet}-Echoes},
	number = {{arXiv}:2405.07354},
    archivePrefix={arXiv},
	publisher = {{arXiv}},
	author = {Gautam, Sushant},
	urldate = {2024-07-19},
	year = {2024},
	archivePrefix={arXiv},
	eprint = {2405.07354 [cs, eess]},
}

@misc{mkhallati_soccernet-caption_2023,
	title = {{SoccerNet}-Caption: Dense Video Captioning for Soccer Broadcasts Commentaries},
	url = {http://arxiv.org/abs/2304.04565},
	shorttitle = {{SoccerNet}-Caption},
	number = {{arXiv}:2304.04565},
	publisher = {{arXiv}},
	author = {Mkhallati, Hassan and Cioppa, Anthony and Giancola, Silvio and Ghanem, Bernard and Van Droogenbroeck, Marc},
	urldate = {2024-07-19},
	year = {2023},
	archivePrefix={arXiv},
    archivePrefix={arXiv},
	eprint = {2304.04565 [cs]},
}

@article{Cioppa2022Scaling,

  title={Scaling up SoccerNet with multi-view spatial localization and re-identification},

  author={Anthony Cioppa and Adrien Deli{\`e}ge},

  journal={Scientific Data},

  year={2022},

  month={June},

  volume={9}

}

@inproceedings{cioppa2022soccernet,
  title={Soccernet-tracking: Multiple object tracking dataset and benchmark in soccer videos},
  author={Cioppa, Anthony and Giancola, Silvio and Deliege, Adrien and Kang, Le and Zhou, Xin and Cheng, Zhiyu and Ghanem, Bernard and Van Droogenbroeck, Marc},
  booktitle={Proceedings of the IEEE/CVF conference on computer vision and pattern recognition},
  pages={3491--3502},
  year={2022}
}

@misc{deepseekai2025deepseekr1incentivizingreasoningcapability,
      title={DeepSeek-R1: Incentivizing Reasoning Capability in LLMs via Reinforcement Learning}, 
      author={DeepSeek-AI and Daya Guo et al.},
      year={2025},
      eprint={2501.12948},
      archivePrefix={arXiv},
      primaryClass={cs.CL},
      url={https://arxiv.org/abs/2501.12948}, 
}

@misc{openai2024gpt4o,
  author       = {OpenAI},
  title        = {Hello GPT-4o},
  year         = {2024},
  url          = {https://openai.com/index/hello-gpt-4o/},
  note         = {Accessed: 2025-04-10}
}

@misc{openai2024introducing,
  author       = {{OpenAI}},
  title        = {Introducing OpenAI o1-preview},
  year         = {2024},
  url          = {https://openai.com/index/introducing-openai-o1-preview/},
  note         = {Accessed: 2025-04-10}
}

@misc{google2023gemini,
  title={Gemini 2.5: Pushing the Frontier with Advanced Reasoning, Multimodality, Long Context, and Next Generation Agentic Capabilities},
  author={{Gemini Team, Google}},
  journal={arXiv preprint arXiv:2507.06261},
  year={2025},
  url={https://arxiv.org/abs/2507.06261}
}

@misc{rao_matchtime_2024,
	title = {{MatchTime}: Towards Automatic Soccer Game Commentary Generation},
	url = {http://arxiv.org/abs/2406.18530},
	shorttitle = {{MatchTime}},
	number = {{arXiv}:2406.18530},
	publisher = {{arXiv}},
	author = {Rao, Jiayuan and Wu, Haoning and Liu, Chang and Xie, Weidi},
	urldate = {2024-07-10},
	year = {2024},
	archivePrefix={arXiv},
    archivePrefix={arXiv},
	eprint = {2406.18530 [cs]},
}

@misc{midoglu_mmsys22_2022,
	title = {{MMSys}'22 Grand Challenge on {AI}-based Video Production for Soccer},
	url = {http://arxiv.org/abs/2202.01031},
	number = {{arXiv}:2202.01031},
	publisher = {{arXiv}},
	author = {Midoglu, Cise and Hicks, Steven A. and Thambawita, Vajira and Kupka, Tomas and Halvorsen, Pål},
	urldate = {2024-12-01},
	year = {2022},
	archivePrefix={arXiv},
	eprint = {2202.01031 [cs]},
}

@misc{rao_towards_2024,
	title = {Towards Universal Soccer Video Understanding},
	url = {http://arxiv.org/abs/2412.01820},
	number = {{arXiv}:2412.01820},
	publisher = {{arXiv}},
    archivePrefix={arXiv},
	author = {Rao, Jiayuan and Wu, Haoning and Jiang, Hao and Zhang, Ya and Wang, Yanfeng and Xie, Weidi},
	urldate = {2024-12-12},
	year = {2024},
	archivePrefix={arXiv},
	eprint = {2412.01820 [cs]},
}

@article{press2022measuring,
  title={Measuring and narrowing the compositionality gap in language models},
  author={Press, Ofir and Zhang, Muru and Min, Sewon and Schmidt, Ludwig and Smith, Noah A and Lewis, Mike},
  journal={arXiv preprint arXiv:2210.03350},
  year={2022}
}

@misc{jiang_active_2023,
	title = {Active Retrieval Augmented Generation},
	url = {http://arxiv.org/abs/2305.06983},
	doi = {10.48550/arXiv.2305.06983},
	number = {{arXiv}:2305.06983},
	publisher = {{arXiv}},
    archivePrefix={arXiv},
	author = {Jiang, Zhengbao and Xu, Frank F. and Gao, Luyu and Sun, Zhiqing and Liu, Qian and Dwivedi-Yu, Jane and Yang, Yiming and Callan, Jamie and Neubig, Graham},
	urldate = {2025-02-12},
	year = {2023},
	archivePrefix={arXiv},
	eprint = {2305.06983 [cs]},
}

@misc{balaji_jersey_2023,
	title = {Jersey Number Recognition using Keyframe Identification from Low-Resolution Broadcast Videos},
	url = {http://arxiv.org/abs/2309.06285},
	doi = {10.48550/arXiv.2309.06285},
	number = {{arXiv}:2309.06285},
    archivePrefix={arXiv},
	publisher = {{arXiv}},
	author = {Balaji, Bavesh and Bright, Jerrin and Prakash, Harish and Chen, Yuhao and Clausi, David A. and Zelek, John},
	urldate = {2025-02-15},
    year = {2023},
	date = {2023-09-12},
	archivePrefix={arXiv},
	eprint = {2309.06285 [cs]},
}

@misc{rao_unified_2020,
	title = {A Unified Framework for Shot Type Classification Based on Subject Centric Lens},
	url = {http://arxiv.org/abs/2008.03548},
	doi = {10.48550/arXiv.2008.03548},
	number = {{arXiv}:2008.03548},
	publisher = {{arXiv}},
	author = {Rao, Anyi and Wang, Jiaze and Xu, Linning and Jiang, Xuekun and Huang, Qingqiu and Zhou, Bolei and Lin, Dahua},
	urldate = {2025-02-18},
	year = {2020},
	archivePrefix={arXiv},
    archivePrefix={arXiv},
	eprint = {2008.03548 [cs]},
}

@misc{sorano_automatic_2020,
	title = {Automatic Pass Annotation from Soccer {VideoStreams} Based on Object Detection and {LSTM}},
	url = {http://arxiv.org/abs/2007.06475},
	doi = {10.48550/arXiv.2007.06475},
	number = {{arXiv}:2007.06475},
	publisher = {{arXiv}},
	author = {Sorano, Danilo and Carrara, Fabio and Cintia, Paolo and Falchi, Fabrizio and Pappalardo, Luca},
	urldate = {2025-02-18},
	year = {2020},
	archivePrefix={arXiv},
	eprint = {2007.06475 [cs]},
}

@article{rafiq2020scene,
  title={Scene classification for sports video summarization using transfer learning},
  author={Rafiq, Muhammad and Rafiq, Ghazala and Agyeman, Rockson and Choi, Gyu Sang and Jin, Seong-Il},
  journal={Sensors},
  volume={20},
  number={6},
  pages={1702},
  year={2020},
  publisher={MDPI}
}

@incollection{kompatsiaris_soccer_2019,
	location = {Cham},
	title = {Soccer Video Event Detection Based on Deep Learning},
	volume = {11296},
	isbn = {978-3-030-05715-2 978-3-030-05716-9},
	url = {http://link.springer.com/10.1007/978-3-030-05716-9_31},
	pages = {377--389},
	booktitle = {{MultiMedia} Modeling},
	publisher = {Springer International Publishing},
	author = {Yu, Junqing and Lei, Aiping and Hu, Yangliu},
	editor = {Kompatsiaris, Ioannis and Huet, Benoit and Mezaris, Vasileios and Gurrin, Cathal and Cheng, Wen-Huang and Vrochidis, Stefanos},
	urldate = {2025-02-18},
	year = {2019},
	langid = {english},
	doi = {10.1007/978-3-030-05716-9_31},
	note = {Series Title: Lecture Notes in Computer Science},
}

@inproceedings{tjondronegoro2003sports,
  title={Sports video summarization using highlights and play-breaks},
  author={Tjondronegoro, Dian and Chen, Yi-Ping Phoebe and Pham, Binh},
  booktitle={Proceedings of the 5th ACM SIGMM international workshop on Multimedia information retrieval},
  pages={201--208},
  year={2003}
}

@article{li_multi-modal_nodate,
	title = {Multi-Modal Large Language Model with {RAG} Strategies in Soccer Commentary Generation},
	author = {Li, Xiang and He, Yangfan and Zu, Shuaishuai and Li, Zhengyang and Shi, Tianyu and Xie, Yiting and Zhang, Kevin},
	langid = {english},
    year = {2025},
}

@article{jin2024bider,
  title={Bider: Bridging knowledge inconsistency for efficient retrieval-augmented llms via key supporting evidence},
  author={Jin, Jiajie and Zhu, Yutao and Zhou, Yujia and Dou, Zhicheng},
  journal={arXiv preprint arXiv:2402.12174},
  year={2024}
}

@article{liang2024kag,
  title={Kag: Boosting llms in professional domains via knowledge augmented generation},
  author={Liang, Lei and Sun, Mengshu and Gui, Zhengke and Zhu, Zhongshu and Jiang, Zhouyu and Zhong, Ling and Qu, Yuan and Zhao, Peilong and Bo, Zhongpu and Yang, Jin and others},
  journal={arXiv preprint arXiv:2409.13731},
  year={2024}
}

@article{khattak2024complex,
  title={Complex video reasoning and robustness evaluation suite for video-lmms},
  author={Khattak, Muhammad Uzair and Naeem, Muhammad Ferjad and Hassan, Jameel and Naseer, Muzammal and Tombari, Federico and Khan, Fahad Shahbaz and Khan, Salman},
  journal={arXiv preprint arXiv:2405.03690},
  volume={1},
  number={3},
  year={2024}
}

@article{shao2024deepseekmath,
  title={Deepseekmath: Pushing the limits of mathematical reasoning in open language models},
  author={Shao, Zhihong and Wang, Peiyi and Zhu, Qihao and Xu, Runxin and Song, Junxiao and Bi, Xiao and Zhang, Haowei and Zhang, Mingchuan and Li, YK and Wu, Y and others},
  journal={arXiv preprint arXiv:2402.03300},
  year={2024}
}

@article{ouyang2022training,
  title={Training language models to follow instructions with human feedback},
  author={Ouyang, Long and Wu, Jeffrey and Jiang, Xu and Almeida, Diogo and Wainwright, Carroll and Mishkin, Pamela and Zhang, Chong and Agarwal, Sandhini and Slama, Katarina and Ray, Alex and others},
  journal={Advances in neural information processing systems},
  volume={35},
  pages={27730--27744},
  year={2022}
}

@misc{zelikman_star_2022,
	title = {{STaR}: Bootstrapping Reasoning With Reasoning},
	url = {http://arxiv.org/abs/2203.14465},
	doi = {10.48550/arXiv.2203.14465},
	shorttitle = {{STaR}},
	number = {{arXiv}:2203.14465},
	publisher = {{arXiv}},
	author = {Zelikman, Eric and Wu, Yuhuai and Mu, Jesse and Goodman, Noah D.},
	urldate = {2025-04-07},
	year = {2022},
	archivePrefix={arXiv},
	eprint = {2203.14465 [cs]},
}

@misc{wang2021knowledgeenhancedsportsgame,
      title={Knowledge Enhanced Sports Game Summarization}, 
      author={Jiaan Wang and Zhixu Li and Tingyi Zhang and Duo Zheng and Jianfeng Qu and An Liu and Lei Zhao and Zhigang Chen},
      year={2021},
      eprint={2111.12535},
      archivePrefix={arXiv},
      primaryClass={cs.CL},
      url={https://arxiv.org/abs/2111.12535}, 
}

@INPROCEEDINGS{wang2025vcass,
  author={Wang, Qixin and Zhou, Songtao and Jin, Zeyu and Guo, Chenglin and Sun, Shikun and Qin, Xiaoyu},
  booktitle={2025 International Joint Conference on Neural Networks (IJCNN)}, 
  title={V-CASS: Vision-context-aware Expressive Speech Synthesis for Enhancing User Understanding of Videos}, 
  year={2025},
  volume={},
  number={},
  pages={1-8},
  doi={10.1109/IJCNN64981.2025.11228511}}

@inproceedings{
jin2026from,
title={From Natural Alignment to Conditional Controllability in Multimodal Dialogue},
author={Zeyu Jin and Songtao Zhou and Haoyu Wang and Minghao Tian and Kaifeng Yun and Zhuo Chen and Xiaoyu Qin and Jia Jia},
booktitle={The Fourteenth International Conference on Learning Representations},
year={2026},
url={https://openreview.net/forum?id=fBagP6w6yE}
}

@inproceedings{zhou2025harmoniVox,
author = {Zhou, Songtao and Qin, Xiaoyu and Zhou, Yixuan and Wang, Qixin and Jin, Zeyu and Wang, Zixuan and Wu, Zhiyong and Jia, Jia},
title = {HarmoniVox: Painting Voices to Match the Avatar's Soul},
year = {2025},
isbn = {9798400720352},
publisher = {Association for Computing Machinery},
address = {New York, NY, USA},
url = {https://doi.org/10.1145/3746027.3755736},
doi = {10.1145/3746027.3755736},
booktitle = {Proceedings of the 33rd ACM International Conference on Multimedia},
pages = {6720–6729},
numpages = {10},
location = {Dublin, Ireland},
series = {MM '25}
}
